\documentclass[preprint]{aastex}
\received{10 August 1998}
\accepted{28 June 2000}

\begin{document}

\title{Exploring Halo Substructure with Giant Stars: 
I. Survey Description and Calibration of the Photometric Search Technique }

\author{Steven R. Majewski\altaffilmark{1,2,3,4}, 
 James C. Ostheimer\altaffilmark{1}, \\
 William E. Kunkel\altaffilmark{5}, and Richard J. Patterson\altaffilmark{1}}

\altaffiltext{1}{Department of Astronomy, University of Virginia,
  Charlottesville, VA, 22903-0818 ({srm4n@didjeridu.astro.virginia.edu,
    jco9w@virginia.edu, ricky@virginia.edu})}

\altaffiltext{2}{Visiting Associate, The Observatories of the Carnegie
  Institution of Washington, 813 Santa Barbara Street, Pasadena, CA
  91101}

\altaffiltext{3}{David and Lucile Packard Foundation Fellow}

\altaffiltext{4}{Cottrell Scholar of the Research Corporation}

\altaffiltext{5}{Las Campanas Observatory, Carnegie Institution of
  Washington, Casilla 601, La Serena, Chile
  ({skunk@roses.ctio.noao.edu)}}

\begin{abstract}
  We have begun a survey of the structure of the Milky Way halo, as well
  as the halos of other Local Group galaxies, as traced by their
  constituent giant stars.  These giant stars are identified via large
  area, CCD photometric campaigns. Here we present the basis for our
  photometric search method, which relies on the gravity sensitivity of
  the Mg I triplet + MgH features near 5150 \AA\ in F-K stars, and which
  is sensed by the flux in the intermediate band $DDO51$ filter.  Our
  technique is a simplified variant of the combined Washington/DDO51
  four filter technique described by Geisler [1984, PASP, 96, 723],
  which we modify for the specific purpose of efficiently identifying
  distant giant stars for follow-up spectroscopic study: We show here
  that for most stars the Washington $T_1-T_2$ color is correlated
  monotonically with the Washington $M-T_2$ color with relatively low
  scatter; for the purposes of our survey, this correlation obviates the
  need to image in the $T_1$ filter, as originally proposed by Geisler.
  
  To calibrate our ($M-T_2$, $M-DDO51$) diagram as a means to
  discriminate field giant stars from nearby dwarfs, we utilize new
  photometry of the main sequences of the open clusters NGC 3680 and NGC
  2477 and the red giant branches of the clusters NGC 3680, Melotte 66
  and $\omega$ Centauri, supplemented with data on field stars, globular
  clusters and open clusters by Doug Geisler and collaborators.  By
  combining the data on stars from different clusters, and by taking
  advantage of the wide abundance spread within $\omega$ Centauri, we
  verify the primary dependence of the $M-DDO51$ color on luminosity,
  and demonstrate the secondary sensitivity to metallicity among giant
  stars. Our empirical results are found to be generally consistent with
  those from analysis of synthetic spectra by Paltoglou \& Bell [1994,
  MNRAS, 268, 793].
  
  Finally, we provide conversion formulae from the ($M$, $M-T_2$) system
  to the ($V$, $V-I$) system, corresponding reddening laws, as well as
  empirical red giant branch curves from $\omega$ Centauri stars for use
  in deriving photometric parallaxes for giant stars of various
  metallicities (but equivalent ages) to those of $\omega$ Centauri
  giants.
\end{abstract}


\section{Introduction}

\subsection{Survey Goals}

Understanding the nature of the Milky Way halo -- its shape, extent,
density distribution, kinematics, abundance distribution and origin --
has long been a central topic in astronomy.  The importance of this
endeavor has increased substantially with the growing, pervasive
connections to a number of other astronomical enterprises bearing on
such wide ranging astrophysical problems as, for example, the magnitude
and distribution of dark matter and the frequency of microlensing
events, the origin of the second parameter problem of horizontal branch
morphology in globular clusters, the nature of high velocity HI clouds,
the interaction of galaxies with their environment, and, of course, the
origin and evolution of the Milky Way.  In spite of the pressing need
for a detailed picture of the halo, at present we still have only the
most rudimentary prescriptions of, for example, the phase space
distribution of halo stars.  Unfortunately, for many applications we no
longer can be satisfied with elementary analytical models of the halo.
Indeed, the very suitability of such simple descriptions may now be
questioned.

A number of new lines of evidence indicate that the halo of the Milky
Way has not achieved a dynamically relaxed state.  Confirmation of this
rather old idea is a long time in coming.  Studies of candidate ``moving
groups'' of metal-poor stars with halo kinematics in the solar
neighborhood were made long ago by Eggen and collaborators
\citep{egg59,egg60}, and the idea of a halo with significant structure
in the form of intermingling ``tube-like swarms'' was a viable
theoretical description at least 35 years ago \citep{oor65}.  However,
apart from consistent attention to the subject of halo ``moving groups''
by \citeauthor{egg60} (\citeyear{egg60,egg77,egg78,egg96}, and
references therein), until recently the subject of halo substructure has
received little other interest, in deference to more prosaic
descriptions of the Galactic halo -- those easily described by
relatively simple analytical prescriptions.  This state of affairs was
influenced perhaps, in part, by a growing emphasis on computer models of
Galactic structure that relied on analytical density laws \citep[ {\it
  et seq.}]{bah81, rob86}, coupled with the lack of {\it systematic}
observational efforts toward uncovering even first order global laws
(e.g., density relations, amount of flattening, kinematical and chemical
trends with position) along more than a small number of lines of sight,
let alone {\it deviations} from these simple global descriptions.
Moreover, the rather smoother phase space distributions of the
``flattened halo'' and Intermediate Population II (which may or may not
be parts of one continuous population; see \citealt{maj93, maj95}),
locally dominate the more extended halo component -- what \citet{oor65}
refers to as the ``pure races of the halo population II''.  Indeed, that
these flattened Galactic populations contain stars with chemodynamical
properties (``high velocity'' and low metallicity) that are typically
considered ``halo-like'' has long vexed understanding of the true
``mixture ratios'' and individual chemodynamical properties of the
various overlapping stellar populations \citep[see, for
example][]{nem93}.  The more relaxed dynamical state of the locally more
dominant, flattened metal-poor components of the Galaxy may have long
distracted attention from an unrelaxed halo population filled with {\it
  substructure}.

With the latter point as preface, it is worthwhile, therefore, to
clarify our own working conception of the ``halo'' -- that Galactic
component we aim to explore in the present survey -- since the very
definition of ``halo'' has taken on such a diversity of connotations.
Presently fashionable ``dual halo'' models of the Milky Way -- those
containing both flattened, prograde rotating and spherical, slow to non-
(or even retrograde) rotating metal-poor components \citep{har87, maj93,
  nor94, car96} -- bear a resemblance to the commonly accepted Galactic
descriptions at the 1957 Vatican Conference \citep{oco58}, when one
appreciates that the properties of the Intermediate Population II (IPII)
discussed there parallel properties assigned to the ``new'' flattened,
contracted halo/thick disk components in today's models.  In the present
survey, we are concerned with Oort's ``pure race'' of Population II: the
extended, more or less spherical\footnote{Even our use of the term
  ``spherical'' here is misleading since it implies a single coherent
  structure with a single density law, whereas if the halo is made up of
  a potpourri of substructured components, the only {\em global}
  coherence of the pure halo races may be in their inexorable response
  to the Galactic potential.} distribution of stars with the most extreme
kinematics in the Galaxy, where growing theoretical and observational
work suggests evidence of past Galactic accretion events may be
fossilized.  The Intermediate Population II, ``low'' or ``flattened''
halo -- which may all be the same thing \citep{maj94} -- we address more
fully in our parallel astrometric survey \citep[e.g.,][ {\it et seq.}]{maj92,cra00}.

In the last decade or so, a number of developments have spawned intense
interest in the interaction of Milky Way-like galaxies with each other
and with their satellites and clusters: e.g., (1) the popularity of Cold
Dark Matter models of the universe, with the concomitant ``bottom up''
growth of structure from the collection of smaller subunits
\citep[e.g.,][]{fre88}, (2) the realization of the prevalence of
gravitationally interacting galactic systems in the nearby universe
(demonstrated of course by \citealt{arp66}; and explored more recently
by, for example, \citealt{mih97}) and at high redshifts by
\citet{van96}, (3) the discovery of the formation of globular clusters
\citep{ash92, sch96} and dwarf satellites \citep*{mir92, hun96} in the
course of such gravitational interactions, (4) the discovery of new
Galactic satellites \citep*{can77,irw90,iba95}, possibly associated
Galactic neighbors \citep*{van91}, and distant globular clusters
\citep*[e.g.,][]{mad82,irw95} and the fact that some of these systems
have complex star formation histories \citep[e.g.,][]{sme94, gre97}, and
(5) the recognition of the complexities of the dynamics of the Local
Group and the implications for the size of galactic dark matter
components and the cosmological density, $\Omega$ \citep{zar89, val93,
  gov97}.  New computer studies of galaxy interactions
(\citealt*{mcg90,moo94,joh96,kro97,kle98}, but all predated by
\citealt{too72}) confirm long-held notions (\citealt{oor65}; see also
the related results of \citealt{inn70}) that satellite galaxies, when
experiencing tidal disruption by a larger galactic potential, can leave
behind long-lived structures -- Oort's ``tube-like swarms'' -- along the
satellite orbit.  The idea that these {\it satellites themselves} may
not be dynamically relaxed systems \citep{kuh89,bur97,kro97,kle98,joh99} has
also played an important role in the debate on the existence of
substantial dark matter halos in these systems \citep[cf. discussion
in][hereafter Paper II]{paperII}.

Hints of substructure in the Galactic halo that may be the ``swarms'' that
are the hallmark of accretion of smaller stellar systems have been
suggested in several surveys of halo stars undertaken since the
concentrated effort of Eggen to find halo (and other) ``moving groups''
among stars in the solar neighborhood.  This more recent 
evidence has most often been found in surveys of the more distant, ``pure''
halo, and most typically as clusterings of stars in distance and radial
velocity -- e.g., the clumps of distant blue horizontal branch stars in
\citet{som87}, \citet{doi89}, and \citet{arn92}.  One very notable radial
velocity-distance clump in the survey of distant stars by \citet{iba95}
is now recognized as the structural paradigm of the type of tidal
events sought here -- an extremely elongated, tidal structure forming
through the destruction of the Sagittarius dwarf galaxy.  Indeed, the
Sgr galaxy was first identified through the clustering of radial
velocities in distant {\it giant stars}, a technique we intend to exploit
in the present survey.

Earlier evidence for possible tidal debris recognized as families of
associated outer halo clusters and dwarf spheroidals was discussed by
\citet{kun76}, \citet{kun79} and \citet{lyn82}, and has received an
increase in interest more recently \citep*{maj94,fus95,lyn95,pal00}.
One ``family'' of clusters (including Arp 2, Terzan 7 and 8, M54, and
Pal 12) is apparently associated with the paradigm tidal event,
Sagittarius \citep{iba95,dm00}.  Phase space clumpings of more nearby,
dwarf stars have also been identified by \citet*{rod90}, \citet*{cot93},
\citeauthor*{mmh94} (\citeyear{mmh94,mmh96}), and \citet*{hel99}, with
the clumps in the latter three works -- delineated in all three dimensions
of motion -- very clearly being identified as halo members on the basis
of both kinematics and abundance.  A surprising implication of
\citet{mmh96} is that the more distant halo appears to be {\it
  dominated} by phase space structure -- i.e., very little
``random'', dynamically relaxed halo field population exists outside of
the IPII.  This empirical result would appear to echo theoretical
suggestions that a large fraction of the halo might contain substructure
\citep{tre93,joh98}.

Detailed surveys of the halo including both precision proper motions and
radial velocities, like the \citet{mmh96} analysis, may be needed to
understand the full spectrum of structure in halo phase space, and that
particular survey continues to pursue that question.  However, to cover
any significant amount of area in this detailed manner presents a
formidable and improbable task at present.  Moreover, to carry out this
kind of work to distances as large as those of the present retinue of
Galactic satellites -- possibly major contributors to the overall halo
field star population -- will need to await new generations of
microarcsecond astrometry instruments, such as the Space Interferometry
Mission.  In the meantime, questions regarding the existence of halo
substructure, its extent, filling factor, size spectrum, etc. may be
approached even when full kinematical data may not be obtainable.  The
survey we describe here is aimed at divining and defining physical
substructure in the halo when that substructure is coherent and has
reasonable contrast above any well-mixed background in the occupied
volume.  A primary aim of the present survey is to probe to large
Galactic distances with ease, and so provide a complement to the more
detailed, but more confined, \citet{mmh96} proper motion-radial velocity
survey of the halo and IPII.  The greater distances probed in the giant
survey described here lend certain advantages over a survey of more
nearby stars, even one replete with full, 3-D kinematical information.
While perhaps decreasing sensitivity to subtle, more diffuse, structures
requiring a full analysis of phase space, exploring at great distances
from the Galactic midplane unfetters the data and analysis from
contamination by stars in the same survey volume having similar
chemodynamical properties to, but coming from stellar populations (e.g.,
IPII, or the ``lower halo'') other than, the ``pure'' halo.  From the
standpoint of uncovering detailed information on the history of the
Galaxy from phase space substructure, surveys of distant halo stars also
have the added benefit that remote tidal structures are longer lived in
the softer gradient of the Galactic potential at large radii.  Debris
closer to the Galactic center becomes phase-mixed rather quickly, and
loses spatial coherence that would otherwise be easily identifiable
(although {\it velocity} coherence is {\it increased} --
\citealt{hw99}).  Distance also increases the contrast of strongly
coherent structures against the background, because the linear width of a
given structure subtends a smaller area on the sky, and the magnitude spread 
in the color-magnitude diagram from depth effects is decreased (see
\citealt{stromlo99}), and because any smooth background of stars should
have a rather steep density fall-off with Galactocentric radius (going
as, say, $r^{-3}$).

\subsection{Survey Approach}

Recent work by \citet{joh96} and \citet{joh98} makes it evident that
kinematic substructure in the halo would be difficult to see with simple
starcounting techniques.  The densities of ``star swarms'' from
dissociated Galactic satellites are simply too low to be detectable
above the foreground curtain of disk and IPII dwarfs -- a problem
especially confounded when the swarms are at distances where the {\it
  much smaller} volume density of their luminous, evolved stars is the
only signal likely to be observable.  To make tidal streams in the outer
halo more evident, it clearly would be beneficial to be able to reduce,
or completely filter out, foreground disk contamination, to which main
sequence stars make the greatest contribution redward of the field main
sequence turnoff.  The signal-to-noise of a systematic search could be
increased substantially if one were able {\it a priori} to key on stars
of a set absolute luminosity because then disk and IPII stars with this
luminosity would be easily differentiated from outer halo stars simply
by their substantially brighter apparent magnitudes.  For example, halo
RR Lyrae stars and horizontal branch stars, which have a relatively
limited range of absolute magnitudes, are rather easy to identify on the
basis of variability or color and have been used as tracers of halo
structure \citep[e.g.,][]{sah85,kin94}.  On the other hand, horizontal
branch stars are sufficiently rare that, while they can point to the
{\it presence} of halo substructure, they may not be able to trace
subtle or more {\it tenuous} halo substructure convincingly and/or
thoroughly, as the rather small, candidate ``moving groups'' found 
by \citet{som87}, \citet{arn92}, and \citet{doi89} would suggest.
Unevolved, main sequence stars would, of course, make up the bulk of any
stellar stream, but it is too difficult at present to explore dwarf
stars to very great distances over large areas, though this technique
has been used to study deep, pencil beam surveys with large telescopes
in small numbers of strategically placed directions \citep[see,
e.g.,][]{gou92,rei96}.

For exploring halo substructure, giant stars may provide a reasonable
compromise between the problems associated with the readily
identifiable, but less numerous horizontal branch stars and the more
common, but intrinsically faint dwarf stars.  Giant stars are generally
a few times more populous than horizontal branch stars in a given
stellar population and provide the distinct advantage that they are
bright enough to be imaged to great distances even with small
telescopes.  With giant stars large volumes of the outer Galaxy may be
explored efficiently with small telescopes (where larger blocks of
observing time are easier to come by) if a reasonable means can be found
by which to pick out the giants from the foreground dwarfs.

There have been several K giant studies at high Galactic latitude
\citep{yos87,rat85,rat89,fly93,mor93} from which we can infer expected
densities of Population II giants. \citet{fly93} found $\sim$ 1 giant
deg$^{-2}$ in a survey complete for the magnitude range
$9.5\lesssim{V}\lesssim{11.0}$ covering $\sim 140$ deg$^2$ at the SGP.
In several high Galactic latitude fields covering a total area of $60$
deg$^2$, \citet{rat85} found a mean density for Population II giants of
order 2 deg$^{-2}$ in the magnitude range $13\lesssim{V}\lesssim{16}$.
With an expected mean radial fall-off of the halo with Galactocentric
radius as $\sim R_{GC}^{-3}$ we obtain a flat differential count of
giants with magnitude (\citefullauthor[see, e.g.,][]{paperII}) and we
may extrapolate that to $V\sim21$ (distances of $\gtrsim250$~kpc, for
typical $M_V\sim-1$ halo metallicity giants) we should expect
approximate mean densities of halo giants of $\lesssim10$~deg$^{-2}$.
Thus, it is clear that a survey for giant stars must cover large areas
of the sky in order to garner reasonably large statistical samples.  The
requirement for large sky coverage drives the criteria to make our
survey as efficient as possible at identifying giant stars.  

It has long been known that the strength of the MgH + Mgb triplet
feature at 5100 \AA\ is strongly dependent on surface gravity \citep[see
Figure 1;][]{ohm34,tha39}, and it is a common technique to identify
giants by their weak absorption in this part of the spectrum
\citep{fri87, iba97}.  \citeauthor{rat83} (\citeyear{rat83, rat85})
showed that discrimination of K giants and dwarfs by this feature was
possible with low resolution (20 \AA\ ) objective prism plates from
Schmidt telescopes, while \citet{mcc76} and \citet{cla79} had already
devised a technique by which to identify giants {\it photometrically}
with a pair of intermediate band filters, centered at 4880 \AA\ ($DDO48$
-- ``continuum'') and 5150 \AA\ ($DDO51$ -- ``Mg'').  This filter system
has been applied to the giant surveys of \citet{yos79,har82,yos87}.

Later, \citeauthor{gei84} (\citeyear{gei84}, {\it et seq.}) showed that
good photometric luminosity classification was still possible with the
$DDO51$ filter (see Figure 1) when the continuum was measured with an
appropriate broad band filter, in this case the $M$ filter of the
Washington system \citep{can76,har79}, at considerable savings in
telescope time. Note that \citet{can76} actually proposed the use of
just such an intermediate band filter located around 5000 \AA\ to allow
luminosity classification. The Washington system itself was designed as
a more efficient broadband system than $UBV$ for the study of the
temperatures and abundances of G and K giants
\citep[see][]{can76,gei86,gcm91}.  Geisler's work has demonstrated the
efficacy of separating G-K dwarfs and giants in the ($T_1-T_2$,
$M-DDO51$) color-color diagram (where $M$, $T_1$, and $T_2$ are in the
Washington system) for a sample of stars spanning a broad range of
metallicity.  Geisler pointed out additional features of his technique
that are beneficial to a survey as described here: (1) The $M-DDO51$
index is insensitive to surface gravity variations among G giants, (2)
the reddening vector in the ($T_1-T_2$, $M-DDO51$) diagram is such that
more reddened, more distant giants will be even {\it more} separated
from less-reddened foreground dwarfs (see Figure 4), and (3) the metallicity
sensitivity of the $M-DDO51$ index is a second order effect, and in the
sense that metal-poor giants have even {\it smaller} Mg absorption.  The
latter feature makes Geisler's system even more useful, since there is
additional discriminating power for our typically expected situation of
selecting metal-poor Population II giants from among foreground
metal-rich dwarfs.

\citet[][ PB94 hereafter]{pal94} demonstrated the gravity discrimination of Geisler's
system using a grid of synthetic spectra over a range of surface
temperatures, gravities and abundances appropriate to both Population I
and II stars, and including realistic sequences of atmospheres for red
giant branch isochrones.  Their work provides a useful demonstration of
the effects of gravity and abundance on the ($M-DDO51$) index: gravity
dominates the color but there is a secondary sensitivity to abundance.
This is illustrated in Figure 2 where the \citeauthor*{pal94} synthetic
colors are translated to loci in Washington/DDO51 color-color planes for
dwarfs and giants of different [Fe/H].

We note here one possible shortcoming of the \citeauthor*{pal94} curves: as
noted by \citet{lej96}, \citeauthor*{pal94} used filter passbands which
differ slightly from the adopted standard Washington filters, and
furthermore, they have used an unpublished grid of model spectra, which
are now somewhat out of date. This probably gives rise to the blueward
translation and small rotation in the color-color plane that we find
necessary in \S 3.6 (Figure 12b and Table 2). \citeauthor*{pal94}
themselves point out that their isochrones are systematically redder
than observed globular cluster giant branches.  Unfortunately,
\citeauthor{lej96} did not include the $DDO51$ filter when they produced
their own (more accurate) synthetic colors. This leaves
\citeauthor*{pal94} as the only source for synthetic photometry with
which we can compare our own photometry. We emphasize that any problems
with the \citeauthor*{pal94} colors appear to be quite small (again, see
\S 3.6), in as much as they relate to our data.

Two important and relevant effects fall out of this interplay of gravity
and abundance as illustrated in Figure 2. The first is that weakened
absorption lines in metal-poor dwarfs that mimic the suppression of Mg
absorption from low gravity in giants becomes a problem only for
subdwarfs with [Fe/H]$\lesssim-2.5$. Main sequence stars more metal-rich
than this, presumably all of those in the disk components and the
majority in the halo as well, are not, in general, confused with giants,
even giants as metal-rich as the Sun. The second relevant effect is
that, for stars on the giant branch, the ($M-DDO51$) color provides a
reasonably good abundance indicator
[$(\partial(M-DDO51)/\partial\rm{[Fe/H]})\sim 0.13$ at ($T_1-T2=0.6$)]
from solar [Fe/H] down to [Fe/H]$\sim-2.0$.

Both of these effects are critical to our enterprise here. First, the
metallicity sensitivity of ($M-DDO51$) we expect to be useful for an
initial sorting of faint giant stars we encounter.  For example, the
appearance of an apparent excess of giant stars of a single
($M-DDO51$)-based abundance would be an expected signal for a tidal tail
from a mono-metallic parent object (a commonly expected paradigm). Such
an excess is clearly seen, for example, in our study of extratidal stars
near the Carina dwarf spheroidal galaxy in \citefullauthor{paperII}
(see, e.g., Figures 7--9 in that paper).

In the case of very metal-poor subdwarfs, we should not encounter an
overwhelming number of contaminants in a sample of giants selected by
($M-DDO51$) techniques, based on the very small fraction of Galactic
stars with metallicities this poor.  From the interim model of
\citet{rei93}, the number of halo intermediate Population II/thick disk
stars expected in the $1\lesssim{M-T_2}\lesssim2$ color range down to
$V=20$ is about 200 deg$^{-2}$ at a Galactic pole. Approximately half of
these stars would be dwarfs and only about 8\% of {\it these} would be
expected to have metallicities [Fe/H] $\lesssim 2.5$, according to
\citet{bee99} and \citet{nor99}. This leaves an expected level of
contamination of $\sim 8$ metal-poor subdwarfs deg$^{-2}$, comparable to
an expected density of giants of $\lesssim 9$ deg$^{-2}$ down to $V=20$.
This is, of course, in the ideal situation of high Galactic latitude. On
the other hand, because of the rapid decline in the metallicity
distribution function below [Fe/H]$\sim-2.0$, it is possible, with only
slightly more conservative giant selection in the color-color plane, to
reduce the subdwarf contamination to practically nothing. For example,
if one were interested in selecting for [Fe/H]$\lesssim-1$ giants, one
would only be concerned with about two [Fe/H]$\lesssim-3$ subdwarfs per
square degree (see Figure 2; \citealt{nor99}).

We believe this level of contamination in our initial photometric
catalogues to be tolerable. In any case, when such stars are
encountered, they will be identifiable by followup spectra (or, for the
brighter stars, by their proper motion from such catalogues as the NLTT;
\citealt*[ {\it et seq.}]{luy79}).  Moreover, these metal-poor stars are
interesting in their own right, and worthy of discovery for further
exploration of the Galaxy's evolution.

For our survey, we have adopted a variant of the Geisler technique that
balances the goals of covering large areas as efficiently as possible
with the desire to obtain as much information about identified giant
stars as possible.  We have adopted a {\it three filter} photometric
system that provides dwarf/giant separation capability, a surface
temperature indicator, and a rough gauge of stellar abundance in giant
stars.  In \citet{gei84}, the $T_1-T_2$ color serves as a surface
temperature index, while the $M-DDO51$ is used as the luminosity index.
However, as we shall show (\S 2.1), the $M-T_2$ color is monotonically
correlated to the $T_1-T_2$ color.  Thus, $M-T_2$ can serve as a
suitable temperature index (as has also been demonstrated by
\citealt*{gcm91}), with the advantage that one less filter is needed in
the observations with no loss in information -- a useful improvement in
efficiency.  In this paper, we explore and calibrate the ($M-T_2$,
$M-DDO51$) diagram for discriminating dwarf and giant stars for a range
of metallicities (\S 2).  

We note that the Washington $C$ filter is designed specifically for
photometric abundance measurement in giant stars, and use of this filter
would provide a much more accurate [Fe/H] for our survey giants
(particularly metal-poor ones) than relying on the secondary dependence
of ($M-DDO51$) on abundance. However, observations in the $C$ filter are
rather expensive (requiring 3 times the exposure time of the $M$ filter
to reach an equivalent depth -- \citealt{can76}).  Since we already
require a large investment in observing time for the $DDO51$ filter, and
our primary imaging goal is to identify giants with as great efficiency
as possible, we have dispensed with using the $C$ filter in most of what
we do. Our rationale is two-fold: (1) We believe that the coarse
abundances afforded by the ($M-DDO51$) color are sufficient for tidal
tail/halo substructure searches in our photometry, and (2) it is our
intention to back up our photometrically identified giant candidates
with spectroscopy as much as possible. Spectroscopy is needed not only
as a check on our giant candidates, but also as a means to obtain
dynamical information from their radial velocities. Much better
abundances may be obtained from moderate resolution spectra (sufficient
for a radial velocity measure) than from $C$ filter photometry. Typical
spectra of the type we are using for this followup work are shown in
Figure 1, where several spectroscopic indicators of surface gravity,
including the Mgb, MgH, and NaD features, may be seen. We will discuss
our spectroscopic work for this survey further in future contributions.

\subsection{Field Selection Strategy}

The selection of fields for our survey reflects two distinct, but
complementary, strategies.  The first is predicated on the paradigm
afforded by the example of the Sgr dwarf galaxy as well as by dynamical
models of tidal effects on satellite galaxies \citep[e.g.,][]
{joh98,joh99}, both which suggest that a non-negligible mass loss rate
in the form of stripped stars may be discernible around presently known
satellite galaxies. Thus, it makes sense to start a search for tidal
tails in the Galactic halo at the most obvious potential sites for their
creation. We have therefore begun a systematic search for giant stars
beyond the tidal radii of the Galactic dwarf satellite galaxies, as well
as a sample of globular clusters. We have already reported successful
searches for extended, coherent stellar structures around our first two
targets, the Magellanic Clouds \citep{stromlo99,victoria98} and the
Carina dwarf spheroidal (\citefullauthor{paperII}).

As the example of the Sgr galaxy illustrates, the tidal debris of
satellite galaxies may stretch to substantial lengths
(\citealt*{mat98,dm00}), and perhaps completely encircle the sky
\citep{joh96,joh98,iba00}. Tracing substantially lengthy tails
continuously outward from the parent could be an extremely
time-consuming enterprise, and should be weighed against the potential
for tracking the path of the debris with more disparate, strategically
placed, pencil beam probes around the sky. Another problem that may be
addressed with a series of probes is the existence of debris from
objects that no longer exist with any recognizable, gravitationally
intact core. In terms of the formation history of the Milky Way halo, it
is obviously of great interest to assess the net contribution of
disrupted bodies to the halo. Finally, in order to understand the
distribution of any ``smooth background'' of halo stars -- the magnitude
of which affects the contrast of any superimposed substructure
\citep{joh96,joh98} -- it makes sense to perform a more systematic
survey that can address the global distribution of giant stars.

For all of these reasons, and others, we have also embarked on the Grid
Giant Star Survey (GGSS). The GGSS is a program to observe 1303
isotropically spaced (mean field spacing $\sim5\fdg6$) Galaxy
pencil-beam probes, each of area 0.4--0.7 deg$^2$, to find giant and HB
stars to explore Galaxy structure. The details and results of this
systematic survey will be presented elsewhere. The ``all-Galaxy'' GGSS
has similar goals and strategy to the ``Spaghetti survey'' described by
\citet{mor00}. This survey also adopts a ($C,M,DDO51,T_2$) photometric
search phase to identify giants and HB stars for spectroscopic followup,
and has presented initial findings and discussion of strategy in
\citet{mor00}. Together, the Spaghetti and GGSS surveys should provide a
wealth of new information on the outer Galaxy.

In \S 2 we determine transformations from our ($M$,
$T_2$) system to both the ($M$, $T_1$, $T_2$) and ($V$, $I$) systems.
In \S 3 we calibrate the ($M-T_2$, $M-DDO51$) color-color diagram for
discriminating dwarfs and giants, and explore the metallicity trends for
giant stars in this diagram. We obtain good agreement in our empirical
calibration of this two-color plane with that given by synthetic
spectra. Finally, by way of $\omega$ Centauri giants as templates, we
determine a rough calibration of giant star absolute magnitudes as a
function of photometric abundance (i.e., position in the two-color
diagram), which can be used for photometric parallaxes.

\section{Photometric Transformations}

\subsection{$T_1-T_2$ to $M-T_2$}

The Washington system was originally designed to obtain temperatures,
metal abundances and CN indices (which relate well to [Fe/H] for
Population I stars, \citealt{jan75}, but  vary independently of
abundance for [Fe/H]$\lesssim{-0.75}$, \citealt{har82}) for G and K giants {\it
  photometrically} \citep{can76}.  The primary goal of the photometric
part of our survey is simpler -- to {\it identify} potential giant
stars.  Thus, whenever possible, candidate giants will be subjected to
spectroscopic follow-up to verify surface gravity (weeding out K
subdwarfs), obtain a radial velocity and determine a {\it spectroscopic}
metallicity\footnote{For those stars where we cannot determine
  spectroscopic abundances, we may fall back on the rough photometric
  metallicity discrimination afforded by the ($M-DDO51$) color.}.  Thus, we
can hope to reduce the number of Washington filter measurements to the
minimum necessary to achieve our survey goals.  For example, at present
we are not emphasizing measurement of carbon abundances, so the CN and G
band-sensitive $C$ filter is not essential to our aims.  On the other
hand, the $M$ filter is needed to serve as a ``continuum'' complement to
the $DDO51$ filter, a \`la \citet{gei84}, and must be retained.

Because the sensitivity of the $M-DDO51$ index to surface gravity is a
function of surface temperature, and because the derivation of
spectroscopic abundances and gravities requires some measure of the
effective temperature, we need an additional photometric measure
sensitive to $T_{eff}$.  In the standard Washington system, the
$T_1-T_2$ parameter serves the role of a temperature index, but for our
purposes greater efficiency could be obtained if a single additional
filter could be combined with the already required $M$ filter to provide
a suitable temperature index.  Unfortunately, the $M$ filter is
subjected to more line blanketing in metal-rich stellar atmospheres than
the $T_1$ filter.  Indeed, in the Washington system \citep{can76}, the
combination of $M-T_1$ with $T_1-T_2$ is intended to provide a
metallicity index, $\Delta(M-T_1)$ (measured from the solar abundance
locus -- similar to the definition of $\delta (U-B)$).  However,
according to Figure 10 of \citep{can76}, the range of $\Delta(M-T_1)$,
is at most about 0.15 magnitudes over $0 \ge$ [Fe/H] $\ge -2.5$ for
giants with a wide range of $T_1-T_2$; and for giants with [Fe/H] $\le
-1.0$ -- typical of the expected metallicities of distant giants we
might hope to find in our survey -- the range of $\Delta(M-T_1)$ is only
about 0.05 magnitudes.  Thus, the $M-T_2$ color may provide an adequate
temperature index for our purposes. Indeed, \citet{lej96} conclude that
``$M-T_2$ is better than $T_1-T_2$ for temperature determinations,
although it is slightly sensitive to surface gravity.''

In Figure 3 we justify our exclusion of $T_1$ imaging in the present
survey as a means to reduce our photometric survey to the simple $M,
T_2, DDO51$ filter system.  Figure 3  shows all stars from Tables V
and VII in \citet[ {\it open squares}]{can76}, Table VI of \citet[ {\it
  solid squares}]{har79}, Table III of \citet[ symbols as in that paper:
{\it open triangles} for luminosity class I-II, {\it solid triangles}
for luminosity class III-IV, and {\it solid circles} for luminosity
class V-VI]{gei84}, Table 2 of \citet[ {\it open circles}]{gei90}, and
solar abundance field giants from Table 1 of \citet[ {\it solid
  triangles}]{gcm91}.  Note, we exclude data for stars from
\citet{can76} when they were reobserved and updated in \citet{har79},
but we include both of the separate measures for the few stars repeated
in \citet{har79} and \citet{gei84}, since these are separate data sets.
Figure 3, which includes dwarfs and giants of a wide range of
metallicities, and which has not been corrected for reddening, already
demonstrates the relative tightness of the monotonic and almost (over
most colors) linear $M-T_2$ versus $T_1-T_2$ correlation.  The upper
dereddening vector in this diagram is from \citet{can76}, while the
lower one is based on the reddening ratios derived in \S 2.3 below.  The
solid line indicates a 4th order fit through the points, ignoring the
two reddest $M-T_2$ stars from \citet{gei90}, which are
in highly reddened, Galactic plane fields.  The root-mean-squared (RMS)
scatter around the fit to this assortment of stars, {\it without}
accounting for differing metallicities and reddening, is only 0.046 (or
0.036 when iteratively excluding greater than 3-$\sigma$ points).

We conclude that while $T_1-T_2$ has been proven an excellent $T_{eff}$
index for G and K stars \citep{can76}, the $M-T_2$ color also provides
an adequate color index suitably correlated to $T_{eff}$ and with only
slight sensitivity to abundances. That we can retain excellent
dwarf-giant discrimination in the {\it three filter} $M, T_2, DDO51$
system is illustrated by the synthetic photometry of \citeauthor*{pal94} in
Figure 2b. In the remainder of this paper we provide an {\it empirical}
formulation for this methodology.

\subsection{Transformation of the ($M$, $T_2$) system to the ($V$, $I$) system}

It is useful to determine the transformation of our Washington ($M$, $T_2$)
filter system into the more commonly used Cousins ($V$, $I$) system.  In 
this way we may compare our field star and cluster color-magnitude data to
the extensive work done in the Cousins system in the literature.  

To determine the transformation functions we take advantage of the
tabulation of photometry by \citet{har79}, who have measured $V$
magnitudes along with Washington photometry in the establishment of
their standard star system.  We also make the assumption that the $T_2$
filter is identical to the Cousins $I_C$ filter.  This assumption is
justified by the great similarity of the $T_2$ bandpass described by
\citet[ see his Figure 1]{can76} and that of the Cousins $I_C$ band as
illustrated by \citet[ see his Figure 1]{bes86}.  Indeed, both bands are
often produced by the same combination of Schott RG-9 filter with
dry-ice cooled Ga-As photomultiplier \citep{bes90}, and similar
effective filter central wavelengths, $\lambda_{eff} \approx 7885$
\AA{,} are determined for giant star colors for the $T_2$ and $I_C$
filters by these two authors.  However, as discussed by Bessell, the
standard stars of \citet{cou81} were established with a photomultiplier
kept $65\arcdeg$C warmer, which gives rise to a 130 \AA\ redward shift
in the $\lambda_{eff}$ in the Cousins standards.  We ignore here any
possible affect this might have on any $T_2$ to $I_C$ transformation.
Presumably, the $T_2$ magnitudes of \citeauthor{har79} would be
identical to the ``natural GaAs'' system $I_C$ band of \citet[ see his
Section III]{bes86}, from which \citeauthor{bes86} notes a slight
non-unity (a slope of 1.036) in conversion to the standard \citet{cou81}
system in $(R-I)$ color, all attributed to the $I$ band shift with
photomultiplier temperature.  Thus, we might expect a systematic error
in the slope of the ($M-T_2$) to $(V-I)_C$ color conversion derived
below of up to 3.6\%.  Also to be kept in mind are any possible
extensions of the redward side of the $I_C$ or $T_2$ response function
for any combination of, say, the RG-9 glass (a ``cut-on'' filter) and
CCD detectors that may have significant response up to the substrate
bandgap cutoff in the near-infrared; this problem would affect the
reddest stars.

For the 79 stars in \citet{har79} that span $-0.285 \le
(M-T_2) \le 2.518$, we obtain a good linear color transformation (RMS
residual = 0.014) as follows:

\begin{equation}
(V-I)_C = (V_C-T_2) = -0.006 + 0.800 (M-T_2).
\label{equation1}
\end{equation}

\noindent A significant contribution to the RMS residual about this fit was 
contributed by the two reddest stars, each having $M-T_2 > 2.02$, in the
sample.  A cubic fit to the data yields no significant improvement.

With the assumption $I_C = T_2$, we find from the above relation that
$V_C$ may be determined by

\begin{equation}
V_C = M - 0.006 - 0.200 (M-T_2).  
\end{equation}

\subsection{Selective Extinction Ratios}

Because the objects we use to calibrate our ($M-T_2$, $M-DDO51$) diagram
in \S 3 are typically at low Galactic latitudes, we must correct
for extinction by dust.  We adopt the average interstellar extinction
curve for $R_V = A_V/E(B-V) = 3.1$ of \citet{sav79}.  For the $M$
($\lambda_{eff}=5085 $ \AA\ , $\lambda^{-1} = 1.97\mu{m}^{-1}$,
\citealt{can76}) and $T_2$ ($\lambda_{eff}=7885$ \AA\ , $\lambda^{-1} =
1.24\mu{m}^{-1}$) filters we obtain $E(M-V) = 0.33 E(B-V)$ and
$E(T_2-V)=-1.27 E(B-V)$.  Thus 

\begin{equation}
E(M-T_2) = 1.60 E(B-V),
\end{equation}

\noindent compared with $E(M-T_2)=1.59 E(B-V)$ from \citet{can79}.  Using older
absorption curves, \citet{can76} derived $E(M-T_2) = 1.67 E(B-V)$.

The effective wavelength of the $DDO51$ filter yields $\lambda^{-1} =
1.94\mu{m}^{-1}$, so that $E(DDO51-V) = 0.27$.  It follows that 

\begin{equation}
E(M-DDO51) = 0.06 E(B-V).
\end{equation}

 We also find 

\begin{equation}
A(M)=3.43 E(B-V).
\end{equation}

\section{Calibration of the Magnesium Index}

In this section, we demonstrate how we discriminate between giants and
dwarfs in the $M-DDO51$ versus $M-T_2$ plane.  Our demonstration
utilizes both previous data in the literature, as well as new
photometric data collected on the Swope 1-m telescope at Las Campanas
Observatory. We also investigate the secondary dependence of the
($M-DDO51$) filter on [Fe/H].

\subsection{Field Star Data from \citet{gei84}}

In Figure 4 we show the $M-DDO51$ versus $M-T_2$ plane with data from
\citet{gei84} and \citet{gcm91} (for solar metallicity field giants),
where photometric data for stars with known luminosity class and
reddening are provided.  Figure 4 here is analogous to Geisler's (1984)
Figure 3.  As can be seen, while there is little ability to discriminate
between supergiants and giants, in general there is excellent
discrimination between dwarfs and evolved stars of all luminosity
classes I-III, as was suggested by the synthetic photometry presented in
Figure 2b.  For $M-T_2 > 1$, subgiants and dwarfs are also well
discriminated, but, as might be expected, the separation begins to break
down for bluer colors, near the main sequence turn off.  With the new
data presented here, we sample this important color regime with a large
number of stars, with the aim of defining a useful guide for dwarf/giant
demarcation in surveys of field stars.

\subsection{New Observations of Star Clusters}

Data for all of the star clusters presented here, except for Melotte 66,
were obtained with the SITe \#1 2048$^2$ CCD attached to the Swope 1-m
on the nights of UT 15-17 March 1997.  All but the very end of the last
night of this run was photometric.  The Melotte 66 data were obtained
with the same filters, telescope and CCD on the photometric night of UT Dec.
15, 1997.

We used the ``Carnegie 3-inch Washington Filter Set'' to which we added a
$DDO51$ filter.  We believe that the Washington filters were purchased
from Omega Optical and that the filters are based on the following
prescription:

\begin{description}
\item {\bf $M$ filter:} 3 mm Schott GG-455 + 5 mm Corning CS-4-96, 
as prescribed by \citet{can76}.
\item {\bf $T_2$ filter:}  9mm Schott RG-9
\item {\bf $DDO51$:} The $DDO51$ filter was purchased from Omega Optical in 
1993, as filter
515BP12, lot \#9349.  This filter is similar to the one described by
\citet{gei84}, but with a slightly larger peak transmission (88\% versus
Geisler's 75\%) and a very slightly narrower passband (half transmission at
about 5095 \AA\ and 5200 \AA\ ).
\end{description}

Our cluster observations were calibrated with Geisler's (1990) SA98,
SA110, and NGC 3680 standard star fields, accounting for color, airmass,
and, in the case of the $M$ band, (airmass) $\times$ (color) terms,
along the lines of the procedure followed in \citet{mkkb94}.  For the
$M$ and $T_2$ solutions, the ($M-T_2$) color was used, whereas, for the
$DDO51$ solution, an ($M-DDO51$) color was used.  For the first two
nights of the March run, during which standard stars were measured, the
solution was allowed to determine individual night zero-points, and
these agreed to better than 0.013 in all cases. The extinction was
assumed to be constant over all three March 1997 nights. In each of the
$M$, $T_2$ and $DDO51$ solutions, there were 88, 69 and 64 useful
standard stars, yielding solutions with RMS errors of 0.0036, 0.0051 and
0.0073 magnitudes, respectively.

\subsection{NGC 3680}

Our photometric catalogue for the intermediate aged \citep{ant91}
cluster NGC 3680 employed ten $M$, eight $T_2$ and nine $DDO51$ frames
having integration times in the ranges of 5-8 seconds, 5-8 seconds and
50-90 seconds respectively.  The images included Geisler's (1984)
calibration sequence in the Washington system.  These CCD frames were
reduced using DAOPHOT II \citep{ste94}, calibrated and checked against
Geisler's sequence, and combined into a single catalogue.  Our CCD
frames cover $23\farcm5$  per side, at $0\farcs68$ per pixel.

In Figure 5 we present the ($M$, $M-T_2$) color-magnitude diagram (CMD)
for 1815 stars in our CCD field of NGC 3680.  Random errors in the
photometry are presented in the right hand panels of Figure 5.  We have
adopted $E(B-V) = 0.046$ (\citealt{nis88}, and consistent with
\citealt*[ NAA97 hereafter]{nor97}), and $A_V = 0.143$, or $E(M-T_2) =
0.074$, $E(M-DDO51) = 0.003$, and $A_M = 0.158$.  \citet{fri93}
determined the abundance of NGC 3680 to be [Fe/H]$=-0.16$, but an analysis
of the CMD based on members cleaned of binary systems by NAA97 gives
[Fe/H] = +0.11 (similar to the value of +0.09 $\pm 0.08$ of
\citealt{nis88}) and an age of 1.45 $\pm$ 0.3 Gyr.

A significant amount of field star contamination is evident in the NGC
3680 CMD, especially around $(M-T_2)_o=1.0$, which is near the main
sequence turn-off of old disk field stars ($(M-T_2)_o\sim0.85$).  The
latter overlap significantly the NGC 3680 main sequence stars of interest.
We have cross-referenced our photometric catalogue to both the proper
motion catalogue of \citet[ K95 hereafter]{koz95} and to the radial
velocity catalogue of NAA97. This allows us to separate true NGC 3680
members from the numerous field star contaminants, and also eliminate
binary stars.  In Figure 6a we show the CMD for only those stars having
measured proper motions by K95 and with determined joint proper
motion-spatial membership probability $P_{\mu,r} > 15\%$.  This low
value was selected to mimic K95's own selection (see Figure 8 in K95)
and intended to preserve some fainter stars in the cluster CMD (faint
stars tend to have lower $P_{\mu,r}$ in the K95 survey).  However, as
K95 point out, this liberal cut probably results in contamination of the
lower main sequence by field stars (e.g., the proper motion errors in
K95 begin to grow substantially for $B>15$).  Therefore, with larger
symbols we denote a more conservative cut with the K95 estimator cutoff
at $P_{\mu,r} > 75\%$.  In Figure 6b we show the resultant
($M-T_2$)$_o$-($M-DDO51$)$_o$ diagram for the proper motion selected
samples.

Use of the much more refined membership analysis of NAA97, which
employed precision radial velocities, clarifies the dwarf/giant
discrimination by colors, albeit only for a bright subsample (Figures 6c
and 6d).  We include the high quality membership data of Figure 6d in
Figure 14, below.  Figure 6c looks qualitatively similar to NAA97's
($b-y, V$) CMD for NGC 3680.  It can be seen, from the general agreement
of the points with the \citeauthor*{pal94} loci for [Fe/H]=0 stars, that
the likely NGC 3680 giants, near $M_o=11.0$, as well as the likely main
sequence stars more or less fall in the appropriate places in the
color-color diagram.

\subsection{NGC 2477}

The rich open cluster NGC 2477 was observed in order to get a large
sample of likely lower main sequence stars.  Unfortunately, extending
the cluster main sequence as faint as possible resulted in the
saturation of our frames for a majority of the suspected cluster giants,
which are lost from our analysis.  We obtained two $M$, three $T_2$ and
two $DDO51$ CCD exposures, for integrations of 60-80 seconds, 60
seconds, and 600 seconds each, respectively.  Again DAOPHOT II was used
to create a calibrated photometric catalogue.  In Figure 7 we present
the ($M$, $M-T_2$)$_o$ CMD for 11,300 stars in the field of NGC 2477.
Previous studies suggest the cluster shows substantial differential
reddening, from $E(B-V)=0.2$ to 0.4 \citep*{har72}; following
\citet{har72} and \citet{smi83}, we have adopted a mean $E(B-V)=0.33$
and $A_V=0.99$.  This translates to $E(M-T_2)=0.53$, $E(M-DDO51)=0.02$
and $A_M=1.13$.  The resultant CMD in Figure 7 is qualitatively
identical with the $(V-I,V)$ CMD of \citet{kas97}, apart from the lack
of a concentration of giants in our CMD.  The broadening of the upper
main sequence has been attributed to the  differential reddening, and
this is supported by the fact that the lower main sequence, which more
nearly parallels the reddening vector, is thinner, particularly in the
($V-I,V$) CMD \citep{kas97}.

Cluster parameters for NGC 2477 have most recently been determined by
isochrone fitting in \citet{kas97}.  This latter study finds best fits
to [Fe/H] = $-0.05 \pm 0.11$ and an age of $1_{-0.2}^{+0.3}$ Gyr.
Earlier age estimates were slightly higher, e.g., 1.5 $\pm$ 0.2 Gyr
\citep{har72} and 1.2 $\pm$ 0.3 Gyr \citep{smi83}, but \citet{car94},
who used the same theoretical isochrones as \citeauthor{kas97} but on
the \citeauthor{har72} data, obtained 0.6 $\pm 0.1$ Gyr.  The white
dwarf cooling curve age of the cluster tends to support the higher ages,
around 1 Gyr \citep*{von95}.  The best fitting \citeauthor{kas97}
abundance for NGC 2477 is slightly lower, but not significantly so, from
the [Fe/H]=0.04 value of \citet{smi83}.

In spite of the amount of attention NGC 2477 has received lately by way
of deep studies \citep{von95,gal96}, no membership study, even at the
bright end of the CMD, exists for this cluster.  In an attempt to
improve our chances of selecting true NGC 2477 main sequence stars so
that we might better delineate the dwarf locus in the ($M-T$, $M-D$)$_o$
diagram, we reduce the field sample by limiting the CMD to stars in two
specific radii -- $7\farcm8$ (Figures 8a-b) and $3\farcm9$ (Figures
8c-d) -- about the nominal cluster center (which was judged by finding
the peak of the X and Y marginal distribution histogram of starcounts
for $M<16$).  By eye we selected stars along the apparently
least-reddened NGC 2477 main sequence and show their position in the
two-color diagram in Figures 8b and 8d ({\it large dots}).  The small
dots show a large population of apparent metal-rich field giants along
the predicted \citeauthor*{pal94} [Fe/H]=0 giant locus, while the predicted
trend for the selected dwarf sequence is clearly borne out by the data.
The spreading of the locus at redder ($M-T_2$)$_o$ colors is a
consequence both of the increased photometric error for the fainter
stars in the sequence as well the contribution of differential
reddening, which increases as the dwarf sequence curves ever more
perpendicular to the reddening vector in the lower part of the diagram.
Note the much larger spread in the unselected stars in Figures 8b and
8d.  This spread about the locus is partly due to large photometric
errors at the faint limit of the catalogue (as high as 0.4 magnitudes in
the colors).  However, the larger bulk of stars {\it above} the dwarf
locus hints at a substantial population of field giants in this low
latitude ($b = -6^o$) field.

\subsection{Melotte 66}

Melotte 66 has long been recognized as an archetypal old open cluster
\citep[e.g.,][]{haw76}.  Recent studies estimate an age for the cluster
of $4.5\pm0.5$ Gyr \citep*{twa95} and $4\pm1$ Gyr \citep{kas97}.
\citet{kas97} adopt an abundance of [Fe/H]=$-0.51\pm0.11$, based on
\citet{fri93}, which is similar to the adopted abundance
([Fe/H]=$-0.53\pm0.08$) of \citet{twa95}.  However, an intrinsic
metallicity spread among the cluster giants has also been detected
\citep{twa95}.  The reddening to the cluster does not seem well
established, although \citet{twa95} settle on $E(B-V)=0.16\pm0.02$ as
the reddening that gives the most consistent ultraviolet excesses
between the cluster dwarfs and giants.  We adopt this value, so that
$E(M-T_2)=0.26$, $E(M-DDO51)=0.01$ and $A_M=0.55$.

We obtained single $M$, $T_2$ and $DDO51$ exposures of Melotte 66 with
integration times of 12, 12 and 100 seconds, respectively, for the Dec
1997 run.  These were calibrated with 67, 59 and 60 useful standard
stars in the $M$, $T_2$ and $DDO51$ bands, respectively, yielding
solutions with random errors of 0.0094, 0.0074 and 0.0109 magnitudes,
respectively.  With these exposure times, we were able to get reliable
photometry for all but the very brightest part of the red giant branch.
We present in Figure 9 the CMD in our filter system for the field of
Melotte 66.  There is substantial field contamination, but the main
sequence turn off region and the red clump of the cluster can be
discerned easily.

While it is in somewhat better shape than either NGC 3680 or NGC 2477,
Melotte 66 is still in need of a better membership census.  Some radial
velocity memberships exist and in Figures 10a and 10b we show those stars
listed as members by \citet{gra82}, \citet{cam87}, \citet{ols91},
\citet{fri93} and from the \citet{geismi84} analysis listed in
\citet{twa95}. The result shows the prominent double CMD sequence caused
by the substantial binary fraction in the cluster.  This duplication
causes both the double subgiant branches as well as a fattening of the
MSTO region (see Figures 9 and 10c).  We show the $(M-T_2, M-DDO51)_o$
diagram corresponding to the identified members in Figure 10b.  Here we
delineate two groups of radial velocity member giants/subgiants: those
falling more or less in the expected region of the color-color diagram
({\it open circles}) and those in more peculiar locations, more like
expectations for dwarf stars ({\it closed circles}).  It can be seen
that the peculiarly-placed stars in Figure 10b seem to be predominantly
associated with the upper subgiant branch, that of the binary stars,
although a few stars, near $(M-T_2)_o \approx 1.1$, are found below this
subgiant branch.  It may be that the wide scatter in $(M-DDO51)_o$ color
near $M-T_2 \approx 1.0$ in Figure 10b (and Figure 10d below), which gives
rise to ``problem giant stars'' in the dwarf region of the two-color
diagram, may be due to problems with the binarity.  It is interesting
also to note that many of these ``problem stars'' have abnormally large
DAOPHOT $\chi$ parameters compared to other stars at similar magnitudes.
Perhaps this is some indication of a slight resolution of close binaries
that is detectable by DAOPHOT.

In Figures 10c and 10d, we attempt to elucidate the true nature of the
red giant branch for the cluster by limiting our field to a radius of
$5\farcm0$.  In Figure 10c we trace likely cluster members using the
known radial velocity members (shown in Figure 10c whether or not they
fall within the $5\farcm0$ radius) and both the single and binary
isochrones in Figure 9d of \citet{kas97} as guides.  Once again, we use
{\it open} and {\it closed} circles to delineate stars with
``problematical'' colors from the standpoint of dwarf/giant separation.
We trace down the red giant branch to the red clump and down the
subgiant branch to the point where both the binaries and the single
stars join their respective main sequence turnoffs.  All other stars
within $5\farcm0$ of the cluster center are left as small symbols in
Figures 10c and 10d.  In Figure 10d the giants above the red clump
follow the \citeauthor*{pal94} locus for solar metallicity giants.  As
expected (from, for example, Figure 4), the subgiant locus in Figure 10d
merges with, and becomes less discriminated from, the locus of main
sequence turnoff stars (small points) at bluer colors.  The latter sweep
lower than the main locus of giant stars in Figure 10d, as expected.
Finally, we note again a predilection for the ``giants'' with
problematical locations in the two-color diagram to be associated with
the subgiant branch for binary members.

\subsection{$\omega$ Centauri}

Because of its proximity and size, $\omega$ Centauri is a convenient
target that provides numerous bright giant stars for use as calibrators.
Moreover, with an abundance ranging from solar to [Fe/H]$< -2.0$
\citep{but78, per80, sun96}, this one cluster allows the opportunity to
explore the abundance sensitivity of the $M-DDO51$ index, as well as to
calibrate the giant star color-magnitude relation as a function of
metallicity for old stellar populations, without introduction of
relative systematic errors produced by differential distance errors.  To
explore the giant distribution in the ($M-T_2$, $M-DDO51$) and the
($M-T_2$, $M$) planes, we have made an extensive imaging survey of the
globular cluster $\omega$ Centauri.

Our $\omega$ Cen data consists of eight separate pointings, offset by
$15\farcm3$ in each the North, South, East, West, Northeast, Northwest,
Southeast, and Southwest directions.  Our resulting catalogue contains
over 100,000 stars photometered, in general, two or three times each,
with DAOPHOT.  The data were matched into catalogues using DAOMASTER.
The Washington CMD for $\omega$ Centauri is presented in Figure 11.  We
have adopted $E(B-V)=0.15$ as in \citet{sun96}; this translates to
$E(M-T_2) = 0.24$, $E(M-DDO51) = 0.009$, and $A_M = 0.51$.  Two points
are worth mentioning about Figure 11.  First, as it is meant only for
illustrative purposes, this figure has not yet been cleaned of multiple
detections on overlapping frames, so that some stars appearing in
overlap regions are represented two to three times.  Second, because of
saturation at the bright end, we have lost some of the very brightest,
reddest $\omega$ Cen giants in this truncated CMD. An interesting aspect
of Figure 11 is the appearance of several distinct RGBs, an aspect we
explore further elsewhere \citep{liege99}.

We have cross-referenced our catalogue with the spectroscopic sample of
\citet{sun96}.  The latter is a careful analysis of abundance patterns
in $\omega$ Cen giants, and also provides a membership list of 343
radial velocity members, stars that were selected in the first place as
likely members based on proper motions.  The stars studied by
\citeauthor{sun96} were in two distinct magnitude ranges, one on the
giant branch and one along the subgiant branch.

In Figure 12a, we show the distribution of the \citeauthor{sun96} giants
in the ($M-T_2$, $M$)$_o$ diagram, and in Figure 12b we show their
distribution in the ($M-T_2$, $M-DDO51$)$_o$ plane.  Giants with
different metallicity ranges are indicated by different symbols.  The
[Fe/H] ranges were selected so that the mean abundance in each
metallicity bin corresponded to more or less convenient [Fe/H] values
(-1.74, -1.60, -1.40, -1.20, -0.95) on the \citet{nor95}
abundance scale; the most metal-rich bin contains a single star at
[Fe/H] = -0.62.  

The distribution of points in Figure 12b demonstrates that while the
$M-DDO51$ index is less sensitive to broad metallicity changes than it
is to luminosity class differences (note the compressed range of
$M-DDO51$ in Figure 12b), the ($M-T_2$, $M-DDO51$)$_o$ plane is still
able to discriminate roughly between metal-poor and metal-rich giants.

The $\omega$ Cen red giant branch is relatively well defined in Figure
11.  A well defined RGB will be useful for our future studies, e.g., as
a means to estimate rough photometric parallaxes for giants with
Washington photometry.  For each abundance, we fit the center of the RGB
sequence analytically (using the IRAF\footnote{IRAF is distributed by
  the National Optical Astronomy Observatories, which are operated by
  the Association of Universities for Research in Astronomy, Inc., under
  cooperative agreement with the National Science Foundation.}task
CURFIT) to equations of the form

\begin{equation}
M_o = a + b(M-T_2)_o + c(M-T_2)_o^2
\end{equation}

\noindent with the results given in Table 1, and plotted in Figure 12a.

To derive analytical descriptions of the rough abundance effect in the
($M-T_2$, $M-DDO51$)$_o$ plane required several steps. We first fit the
various metallicity bins with relations of the form

\begin{equation}
(M-DDO51)_o = d + e(M-T_2)_o + f(M-T_2)_o^2.
\end{equation}

\noindent Especially at the extreme ends of the $\omega$ Cen abundance range,
however, we found these fits to be unsatisfactorily constrained due to
the small sample sizes, poor distribution in $(M-T_2)$ color, and the
contribution of photometric errors, which become larger than the mean
separation of $(M-DDO51)$ colors for the metal-poor loci. The bimodal
$(M-T_2)$ distribution of the data gave rise to unlikely ``bowing''
between nodes fixed by the two concentration of points. Therefore, to
give a more realistic and natural fit to the data, we have taken
advantage of the synthetic curves derived for cluster CMDs by
\citeauthor*{pal94}, which we ``calibrate'' to our data for $\omega$ Cen. This
forging of the synthetic loci to the actual data has the added advantage
of extending the usefulness of the $\omega$ Cen example to abundance
ranges outside those encompassed by the cluster itself, if we assume the
{\it relative} placement of the synthetic loci is accurate. After trying
various schemes to marry the synthetic and empirical loci, we found that
we could obtain reasonable matches of the \citeauthor*{pal94} curves to
the empirical distributions by metallicity with a simple 0.25 mag blueward
translation (in $M-T_2$) and a slight ($<2\arcdeg$) rotation of the
\citeauthor*{pal94} curves in the two-color plane. Thus we find
\begin{equation}
(M-T_2)_o=(M-T_2)_o^{PB94} - 0.25,
\end{equation}

\noindent and,

\begin{equation}
(M-DDO51)_o=(M-DDO51)_o^{PB94}-0.029*(M-T_2)+0.03.
\end{equation}

\noindent Note these relations are defined for the \citeauthor*{pal94} RGB 14 Gyr 
isochrones, and may not be applicable to their other models, and for
clarity they were not considered in the previous discussion of the open
clusters.

As pointed out in \S 1.2, the need to transform the \citeauthor*{pal94}
relations is likely due to the nonstandard passbands adopted by
\citeauthor*{pal94} \citep{lej96}, and it confirms the finding by
\citeauthor*{pal94} that their isochrones are systematically redder than
actual globular cluster red giant branches.  Table 2 includes the
original empirical fits to the metallicity bins in Table 1. We include
only the [Fe/H]$=-1.74$, $1.40$ and $-1.20$ groups, which guided our
transformation, in Figure 12b (dashed lines), along with the revised
\citeauthor*{pal94} RGB loci (solid lines), and illustrates the
reasonable matches of the latter to the $\omega$ Cen data. For future
analytical ease the rotated \citeauthor*{pal94} loci have been fit by
equations (7).  The coefficients for these new relations are given in
Table 2 for the various metallicities.  The difference in the mean
abundances for the listed loci in Table 2 compared to Table 1 is a
result of our adopting the \citeauthor*{pal94} RGB curves (and therefore
their abundance selections) for the latter.

To obtain the absolute magnitude, $M_M$ for a star using the relations
in Table 1, one must correct the calculated apparent magnitude for the
$\omega$ Cen distance modulus, which may be taken as (m-M)$_o =13.57$
\citep{dic88}, as well as the extinction in the $M$ band ($A_M = 0.51$
for $E(B-V)=0.15$).  It should be pointed out that since we have not
been able to discriminate between first ascent red giant branch and
asymptotic giant branch stars, the relations in Table 1 may tend to give
systematically overluminous magnitudes if applied to individual first
ascent giant branch stars.  The ratio of first ascent to asymptotic
giant branch stars in the brighter magnitude range in Figure 12a (i.e.
above the horizontal branch) may be something like four to one
\citep{sun96}.  On the other hand, if the halo of the Galaxy evolved
roughly similarly to $\omega$ Cen, with a similar age-metallicity
relation and initial mass function, then we might expect a similar first
ascent to asymptotic giant branch ratio in the field, and application of
the Table 1 relations to finding distances for halo field stars should
provide results that are at least statistically correct for an ensemble
of giant stars.

\subsection{Previously Published  Data}

In a series of papers over more than a decade Doug Geisler and
collaborators (\citealt{gei86, gei87, gei88}, \citealt*{gcm91, gcm92,
  gcm97}, \citealt*{gmc92}), have continued to refine calibration of the
Washington photometry system through extensive observations of giant
stars in open and globular clusters of all metallicities.  In many
cases, the Washington data presented have been supplemented with
observations in the $DDO51$ filter as a means to discriminate cluster
giants from foreground dwarfs.  In Figure 13b we compile these data in
the two-color diagram of interest to our analysis (similar to Figure
12b).  Many of the stars were observed less frequently in $DDO51$ than
in the other filters, and this has resulted in larger errors in the
$M-DDO51$ color.  Because of the sensitivity of the technique on
relatively reliable $M-DDO51$ colors, we have excluded all stars which
have fewer than three measures in $DDO51$.  In Figure 13a we show the
$(M-T_2, M_M)$ color-absolute magnitude diagram for those stars with
available magnitude data in addition to colors.

Table 3 summarizes the data employed in the construction of Figure 13.
In both Table 3 and Figure 13 the clusters are grouped according to
[Fe/H] (with groups defined by the curves in Figure 13b), where the
[Fe/H] and other data are from \citet{mer98} for the open clusters and
from \citet{har98} for the globular clusters, with two exceptions: the
distance moduli for NGC 2360 and NGC 6809 are taken from \citet{mer90}
and \citet{alc92}, respectively. For these particular clusters the
values listed in the databases of \citet{mer98} and \citet{har98} differ
significantly from those in the literature (and we suspect an error in
the databases).  Figure 13 demonstrates that the general trends with
[Fe/H] seen in the $\omega$ Cen giants are exhibited in the cluster
giant population as a whole -- namely, that the giants in the most
metal-poor clusters show almost no variation of $(M-DDO51)$ color with
$(M-T_2)$, while more metal-rich cluster giants have progressively
smaller $(M-DDO51)$, particularly at redder $(M-T_2)$, as shown in
Figure 12b.  We have plotted the five color-color curves calibrated by
the various metallicity populations in $\omega$ Cen (Table 2), with
metallicity increasing towards the curves with smaller $(M-DDO51)$. We
also include in Figure 13b the rough dwarf/giant demarcation ({\it
  straight solid lines}) recommended in the next Section.  It may be
seen that, apart from the very reddest and the very bluest stars, the
majority of the giants lie above this demarcation line.  The red upturn
in the demarcating line is necessitated by the overlap with red dwarfs
at $(M-T_2)>2.0$ (see Figure 14) .

For a given abundance range the cluster data in Table 3, however, show
much larger scatter in both Figure 13a and 13b compared to the $\omega$
Cen giants in Figures 12a and 12b.  This is partly due to the
heterogeneity in the sources and techniques employed to measure the
fundamental parameters of distance, abundance and reddening for the
Table 3 data.  Of course, there is also a considerable spread in ages in
the objects in Table 3, particularly with the inclusion of the open
clusters, which contributes to the spreading of the distributions in
Figure 13. In contrast, the \citeauthor*{pal94} isochrones we have used
were derived from models with ages (14 Gyr) appropriate for globular
cluster giant branches.

Because of the various problems with the heterogeneity of the data, we
do not use Figure 13 to calculate general formulae for the conversion of
color into absolute magnitude, as we did in Tables 1 and 2 for the
$\omega$ Cen giants.  We provide the Figure 13 data here to lend
ancillary support to our assertion that rough abundance discrimination
is possible in the $(M-T_2, M-DDO51)$ diagram.  While it would be
nice to extend our Table 1 calibrations to higher abundances through
inclusion of the metal-rich open clusters, for our present purposes
there is no urgency for this since at the magnitudes and Galactic
latitudes of our first applications of the technique we do not expect to
encounter many giants with the metallicities and ages in the range of
the metal-rich open clusters.  We aim to return to this problem if and
as needed.

We reiterate that much better photometric abundances can be had with
inclusion of the Washington $C$ filter, as the numerous references
listed in Table 3 can attest.  However, our goal here is (1) to identify
the minimal amount of photometric data necessary to {\it find} field
giants that we have already begun to observe spectroscopically for
radial velocities and, of course, abundances, and (2) to point out the
additional leverage that our three color filter system has in
characterizing the relative abundance of our giant star candidates.

\subsection{Giant-Dwarf Discrimination}

In Figure 14, we combine the two-color data of our open cluster main
sequences from NGC 3680 (both single and binary radial velocity members
in Figure 6d) and NGC 2477 (those stars selected as main sequence in
Figure 8d), our red giants from NGC 3680 (both single and binary radial
velocity members in Figure 6d) and $\omega$ Cen (Figure 12b), the red
giant and subgiant branch of Melotte 66 (radial velocity members from
the single star locus only, Figure 10b), the solar abundance field
giants from \citet{gcm91}, as well as the data on field dwarfs and
giants from \citet{gei84}, and giants from \citet{gcm91} (Figure 4) and
the clusters listed in Table 3.  All evolved stars, from Geisler's
single luminosity class ``IV-V'' to class I, are lumped into one group
(``non-dwarfs''), as are all of the stars we show from the Melotte 66
red giant/subgiant branch.  Note that \citet{gei84} specifically
investigates the sensitivity of the $(M-DDO51)$ color to giants of
different surface gravities and he finds little separation for giants of
similar temperatures and metallicities but different log$g$.

We see that the discrimination of evolved stars and dwarfs is
rather good over a broad range of $M-T_2$, breaking down only at the 
bluish colors ($M-T_2 \approx 0.7-0.9$) of subgiant stars just evolved 
from the main sequence, and at the red end, starting around the colors of 
late K stars ($M-T_2 \approx 2.3$), where overlap with dwarfs occurs again.  
The {\it straight solid lines} in the diagram roughly demarcate the region
above which one might expect to find predominantly subgiant and
giant stars.  The ``non-dwarf'' objects that fall below these lines in 
Figure 14 are generally Geisler stars classified as subgiants, the 
lowest luminosity Melotte 66 subgiants, as well as the peculiar group
of Melotte 66 stars (with $1.0 \le (M-T_2)_o \le 1.2$) discussed above. 

The rather clear separation in Figure 14 motivates our survey for distant
giant stars.  With a 1-m class telescope and CCD imaging it is easy to
obtain the photometric precision necessary to pick out rather complete
samples of giant stars to distances of hundreds of kiloparsecs.  By
adopting this technique, we intend to explore the halo and local
environment of the Milky Way and other galaxies in the Local Group with
modest sized telescopes.

\acknowledgments We greatly appreciate the assistance of Wojtek
Krzemi\'nski and Beata Mazur who collected the March 1997 CCD data
presented in this paper while WEK worked at another telescope.  We thank
the referee, Gordon Drukier, for comments that helped to improve the
text. We gratefully acknowledge many helpful discussions with and
comments from Doug Geisler, and thank him for providing some of his data
in machine readable form.  We also appreciate helpful discussions about
(and catalogued data for) $\omega$ Centauri with Noella D'Cruz, Robert
O'Connell and Robert Rood.  JCO thanks Ata Sarajedini for helpful
discussions and mentoring during JCO's 1997 Research Experience for
Undergraduates stay at Kitt Peak National Observatories, when part of
this work was done.  JCO and SRM acknowledge several grants in aid of
undergraduate research from the Dean of the College of Arts \& Sciences
at the University of Virginia.  SRM also acknowledges partial support
from an National Science Foundation CAREER Award grant, AST-9702521, a
fellowship from the David and Lucile Packard Foundation, and from a
Cottrell Scholarship from the Research Corporation.

SRM and WEK dedicate this work to the memory of their 
friend and colleague, Jerry Kristian.

\begin{deluxetable}{cccccccc}
\tablewidth{0pt}
\scriptsize
\tablecaption{Fits to $\omega$ Cen giant branches in the $(M-T_2, M)_o$ color-magnitude diagram.}
\tablehead{\colhead{[Fe/H]}      & \colhead{$\langle{\rm [Fe/H]}\rangle$} & \colhead{\#stars} & \colhead{$(M-T_2)_o$} & \colhead{$a$\tablenotemark{a}}      & \colhead{$b$\tablenotemark{a}}       & \colhead{$c$\tablenotemark{a}}      & \colhead{RMS}}
\startdata
$-$0.80 to $-$1.05 & $-$0.95     & 10     & 1.14-2.05   & 25.3408 & $-$12.2239 & 2.75723 & 0.242 (10)  \\
$-$1.06 to $-$1.29 & $-$1.20     & 18     & 1.04-1.95   & 25.9265 & $-$13.9932 & 3.46823 & 0.236 (17)  \\
$-$1.30 to $-$1.47 & $-$1.40     & 31     & 1.01-1.69   & 27.6675 & $-$16.5382 & 4.20810 & 0.196 (31)  \\
$-$1.48 to $-$1.67 & $-$1.60     & 76     & 0.98-1.85   & 29.5291 & $-$19.3737 & 5.18975 & 0.231 (76)  \\
$-$1.68 to $-$1.93 & $-$1.74     & 91     & 0.85-1.94   & 31.7516 & $-$22.9986 & 6.52318 & 0.206 (88)  \\
\enddata
\tablenotetext{a}{Coefficients from equation 6 (see text).}
\end{deluxetable}

\begin{deluxetable}{cccc}
\tablewidth{0pt}
\scriptsize

\tablecaption{Fits to $\omega$ Cen giant stars in the $(M-T_2,
  M-DDO51)_o$ color-color plane. }

\tablehead{ \colhead{[Fe/H]} &  \colhead{$d$\tablenotemark{a}}      & \colhead{$e$\tablenotemark{a}}       & \colhead{$f$\tablenotemark{a}}       }
\startdata
\sidehead{\it empirical curves}
 $-$1.20     & $-$0.02598 &\phs0.12815 & $-$0.07806\\
 $-$1.40     & $-$0.14206 &\phs0.31176 & $-$0.13879\\
 $-$1.74     &\phs0.03193 &\phs0.02258 & $-$0.01605\\
\sidehead{\it synthesized curves\tablenotemark{b}}
 $-$0.47     & $-$0.12560 &\phs0.35652 & $-$0.23447 \\
 $-$0.79     & $-$0.08477 &\phs0.27546 & $-$0.17658 \\
 $-$1.25     & $-$0.05076 &\phs0.18657 & $-$0.10393 \\
 $-$1.77     &\phs0.01888 &\phs0.04394 & $-$0.02363 \\
 $-$2.03     &\phs0.02907 &\phs0.02549 & $-$0.01225 \\
 $-$2.23     &\phs0.02336 &\phs0.03606 & $-$0.01415 \\
 
 \enddata 
\tablenotetext{a}{Coefficients from equation 7 (see text).}
\tablenotetext{b}{Fits for the synthesized curves are
   constrained by the color-color relations in \citeauthor*{pal94} (see
   text).}  
\end{deluxetable}

\begin{deluxetable}{lrlrrrlll}
\tablewidth{0pt}
\tabletypesize{\scriptsize}
\tablecolumns{9}

\tablecaption{Star clusters used in the construction of Figure 13.}

\tablehead{\multicolumn{2}{c}{Cluster} &\colhead{}& 
\multicolumn{3}{c}{Global Cluster Data} &
\colhead{}&\multicolumn{2}{c}{Photometry Data} \\
\cline{1-2} \cline{4-6} \cline{8-9}
\colhead{Name} &\colhead{Type} &\colhead{}&
\colhead{[Fe/H]\tablenotemark{a}}
&\colhead{$E_{B-V}$\tablenotemark{a}}
&\colhead{$(m-M)_V$\tablenotemark{a}} 
&\colhead{}
&\colhead{stars\tablenotemark{b}}
&\colhead{reference\tablenotemark{c}}}

\startdata 
NGC 7099    &globular&&$-2.12$ &0.03   & 14.57 && 14(6) & G88,GMC92\\
&&&&&&&&\\
NGC 6809    &globular&&$-1.81$ &0.07   & 14.10\tablenotemark{e} && 15(6) & GMC92    \\
&&&&&&&&\\
NGC 6752    &globular&&$-1.55$ &0.04   & 13.08 && 22(3) & GCM97    \\
NGC 1904    &globular&&$-1.54$ &0.01   & 15.53 && 11(1) & GCM97    \\
NGC 1261    &globular&&$-1.35$ &0.01   & 16.05 &&  7(5) & G88    \\
NGC 6121    &globular&&$-1.28$ &0.36   & 12.78 && 23(12)\tablenotemark{d} & G86  \\
NGC 1851    &globular&&$-1.26$ &0.02   & 15.49 &&  9(6) & GCM97    \\
&&&&&&&&\\
NGC 362     &globular&&$-1.16$ &0.05   & 14.75 && 10(10)\tablenotemark{d} & G86  \\
NGC 6723    &globular&&$-1.12$ &0.05   & 14.82 && 13(13)\tablenotemark{d} & G86  \\
&&&&&&&&\\
NGC 6637    &globular&&$-0.71$ &0.17   & 15.11 &&  9(7)\tablenotemark{d} & G86  \\
NGC 6352    &globular&&$-0.70$ &0.21   & 14.39 && 19(1) & G86       \\
&&&&&&&&\\
NGC 2243    &open&&$-0.44$ &0.03   & 13.17 &&  6(6) & G87      \\
NGC 2204    &open&&$-0.34$ &0.08   & 13.44 &&  8(8) & G87      \\
Melotte 71 &open&&$-0.29$ &0.01   & 12.53 &&  9(7) & GCM92    \\
NGC 2360    &open&&$-0.15$ &0.09   & 10.30\tablenotemark{f} &&  6(6) & GCM92    \\
NGC 6940    &open&&$+0.01$ &0.23   & 10.83 && 10(10)\tablenotemark{d} & C86  \\
NGC 6705    &open&&$+0.14$ &0.42   & 12.61 && 22(6)\tablenotemark{d} & C86  \\
NGC 6791    &open&&$+0.15$ &0.18   & 13.93 && 12(2)\tablenotemark{d} & C86  \\
\enddata
\tablenotetext{a}{[Fe/H], reddening and apparent visual distance modulus
  from \citet{mer98} for open clusters and \citet{har98} for globulars.
  Note that $(m-M)_V=(m-M)_M + E(m_M-m_V)$.}

\tablenotetext{b}{Number of stars with $(M-DDO51)$ and $(M-T_2)$ data.
  The number in parentheses is the number of stars with at least three
  observations in $(M-DDO51)$. Only these stars are included in Figure
  13. Note: stars excluded by GCM92 as nonmembers were excluded here
  also. }

\tablenotetext{c}{Reference (listed below) for $(M-DDO51)$, $(M-T_2)$,
  and, when available, $T_1$ magnitudes (for calculation of $M$
  magnitudes from $M-T_1$ colors).}
\tablenotetext{d}{Indicates that no $T_1$ data were available to compute
  $M$ magnitudes from the listed $(M-T_1)$ colors, so these objects do
  not contribute to Figure 13a, but are included in Figure 13b.}
\tablenotetext{e}{Distance Modulus from \citet{alc92}}
\tablenotetext{f}{Distance Modulus from \citet{mer90}}
\tablerefs{
C86 -- \citet{can86}; 
G86 -- \citet{gei86}; 
G87 -- \citet{gei87}; 
G88 -- \citet{gei88}; 
GCM92 -- \citet*{gcm92};
GCM97 -- \citet*{gcm97} -- stars classed as dwarfs left out; 
GMC92 -- \citet*{gmc92}}

\end{deluxetable}

\clearpage 
\begin{figure}
\includegraphics[angle=0,scale=1]{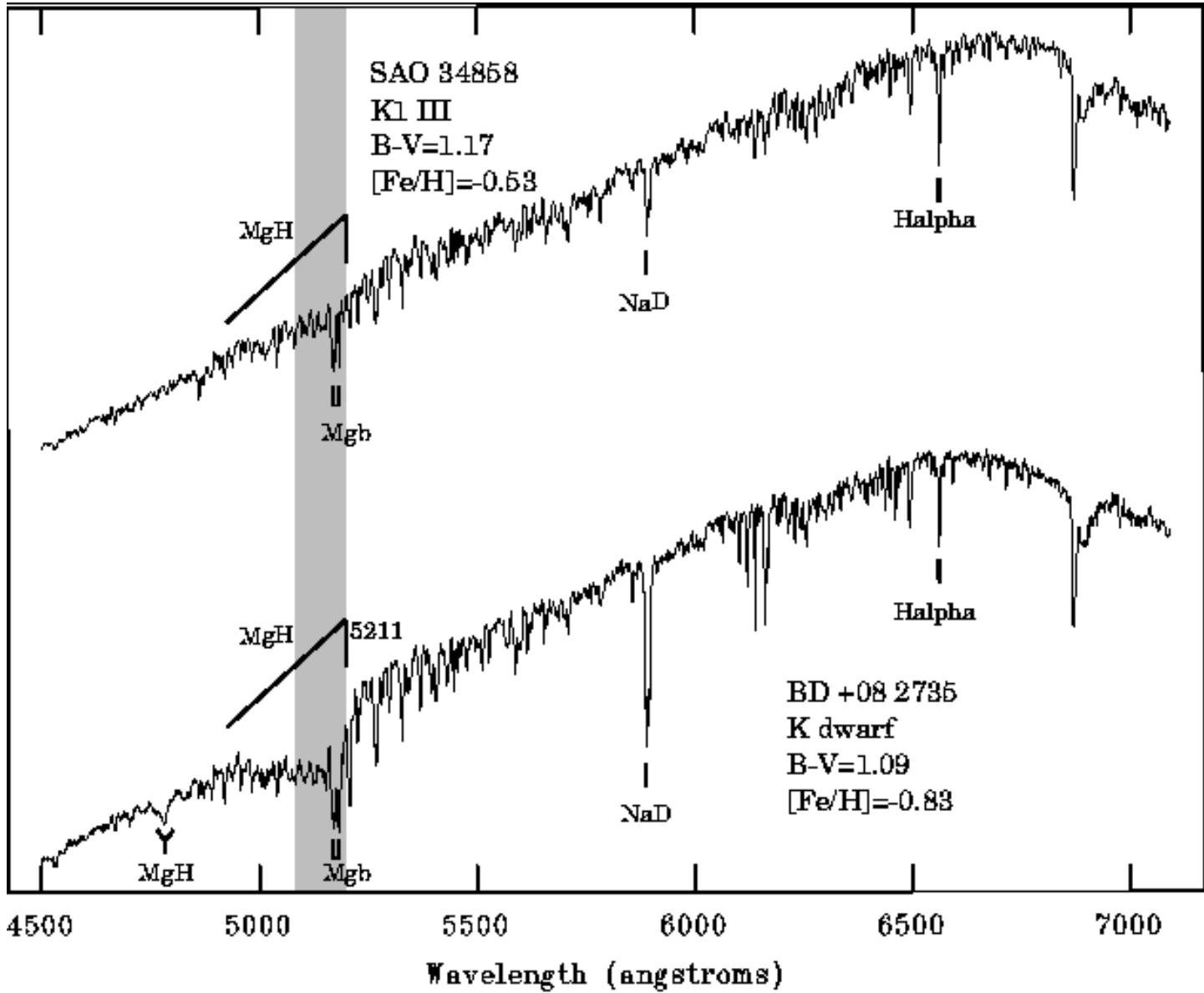}
\figcaption[1.ps]{Comparison of spectra for K giant and dwarf stars
  of similar color and abundance, illustrating the dependence of the MgH
  + Mgb triplet on luminosity class. The location of the $DDO51$ filter
  bandpass is indicated by the shaded region. Note also the
  gravity-sensitivity of the MgH band near 4850 \AA\ as well as the NaD
  doublet \citep{tri97}.}
\end{figure}

\begin{figure}
\includegraphics[angle=0,scale=1]{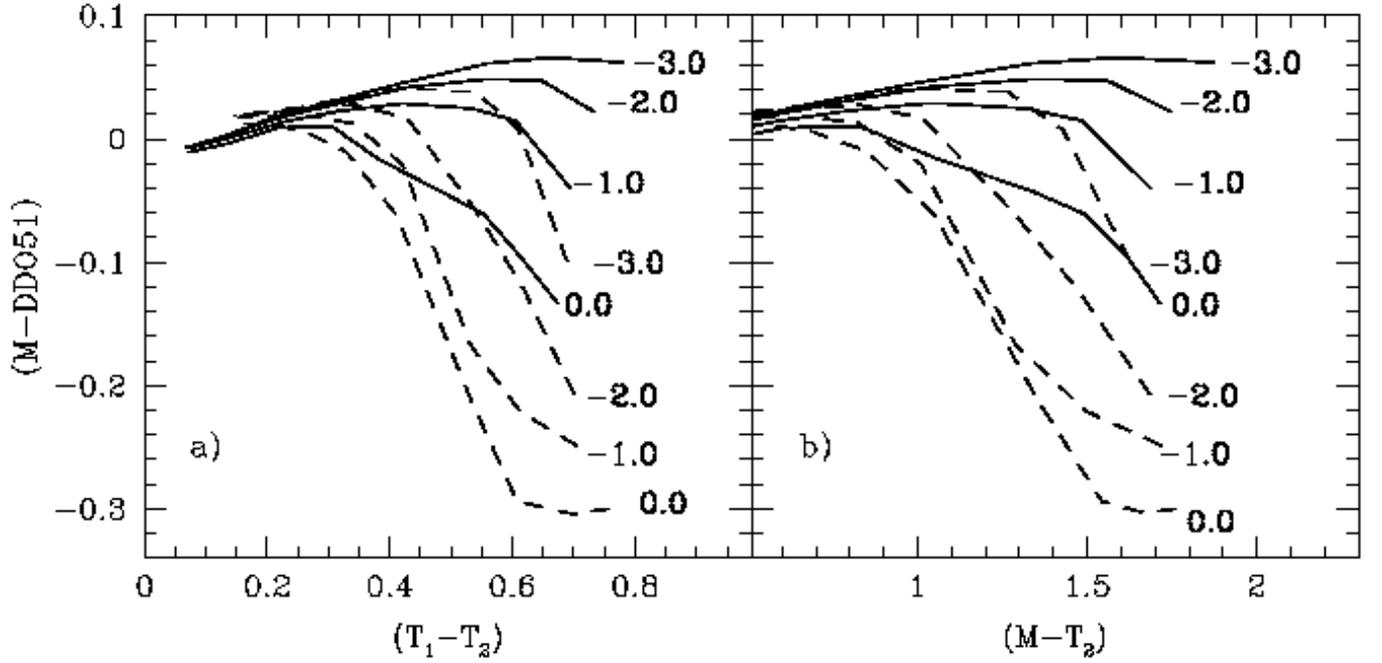}
\figcaption[2.ps]{(a) The $(T_1-T_2, M-DDO51)$ two-color diagram as
  invented by \citet{gei84}, illustrated with the loci for giants {\it
    solid lines} and dwarfs {\it dashed lines} for the indicated values
  of [Fe/H], derived from the synthetic spectra generated by
  \citeauthor*{pal94}. (b) Same as (a) but for the $(M-T_2, M-DDO51)$
  two-color plane adopted for our survey.}
\end{figure}

\begin{figure}
\includegraphics[angle=0,scale=1]{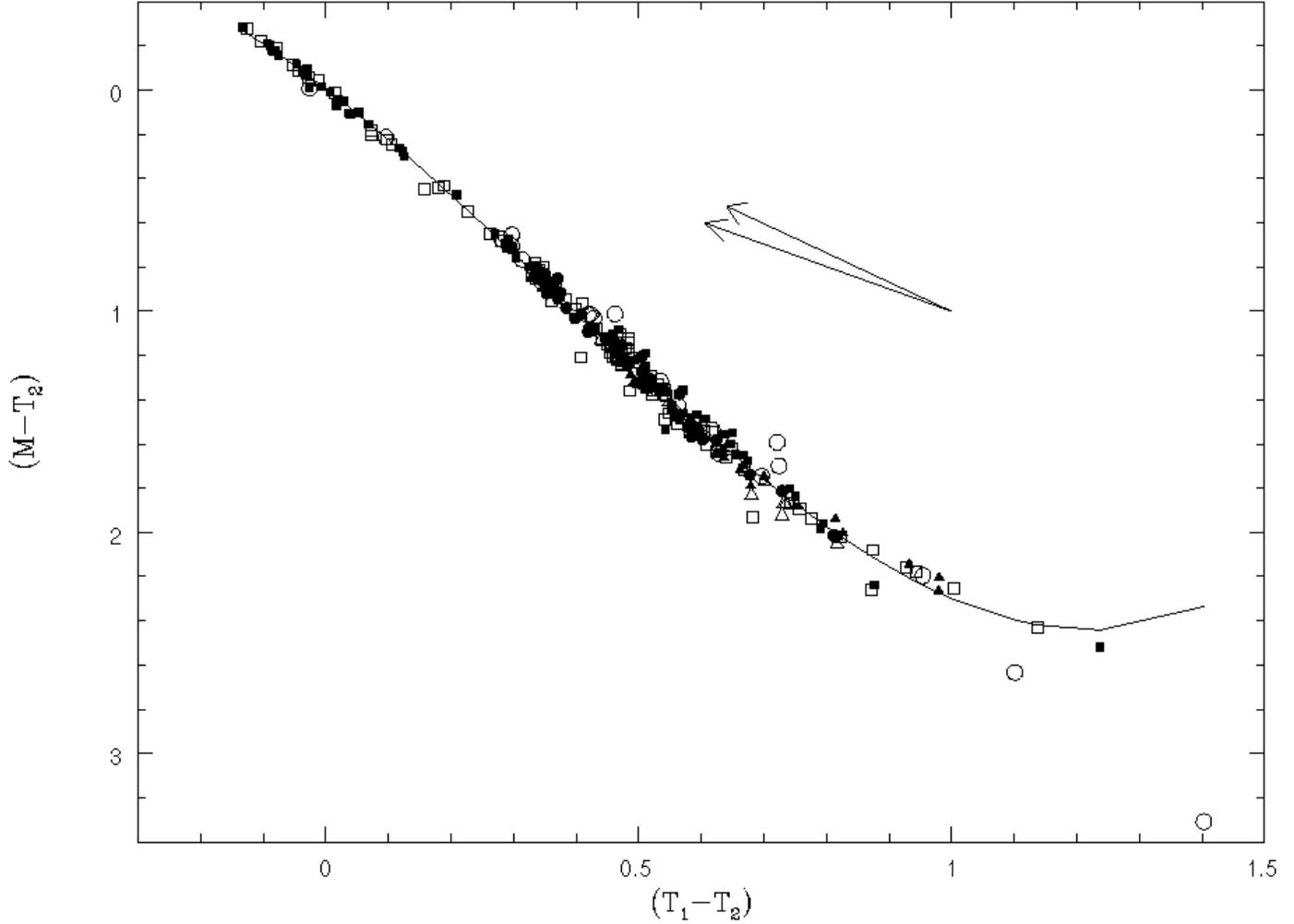}
\figcaption[3.ps]{The $(M-T_2)_o$ versus $(T_1-T_2)_o$
  plane using data from the sources cited in the text.  Symbols denote
  luminosity classes, according to the legend in Figure 4.  The upper
  reddening vector is from \citet{can76}, the lower one is taken from
  the analysis in \S 2.3.  The {\it solid line} is a 4th order fit to
  the data, apart from the reddest two data points.  The data include
  dwarfs and giants from a wide range of metallicities, is not corrected
  for reddening, and yet shows a relatively tight correlation of $M-T_2$
  to $T_1-T_2$.}
\end{figure}

\begin{figure}
\includegraphics[angle=0,scale=1]{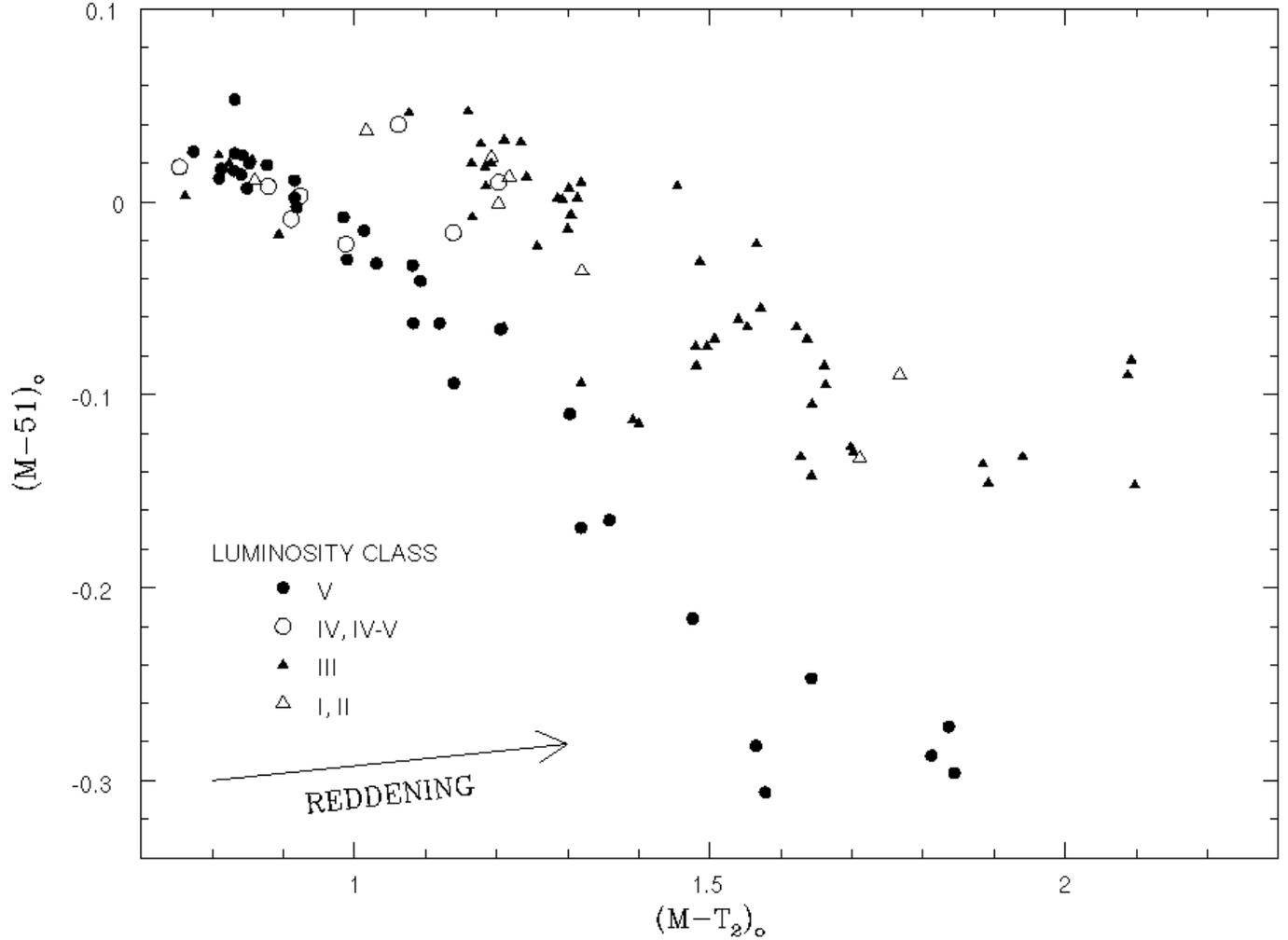}
\figcaption[4.ps]{The $(M-DDO51)_o$ versus $(M-T_2)_o$
  plane using data from field stars from \citet{gei84} and solar
  abundance field giants from \citet{gcm91}.  Symbols denote luminosity
  classes as given in the legend.  Figure 4 here is analogous to
  Geisler's (1984) Figure 3, which shows the same general trends with
  luminosity class and color in the $(M-DDO51)_o$ versus $(T_1-T_2)_o$
  plane.  While there is little ability to discriminate between
  supergiants and giants in our two color plane, there is excellent
  discrimination between dwarfs and evolved stars of luminosity classes
  I-III redward of $M-T_2$ colors typical of the main sequence turn off
  for old populations.}
\end{figure}

\begin{figure}
\includegraphics[angle=0,scale=1]{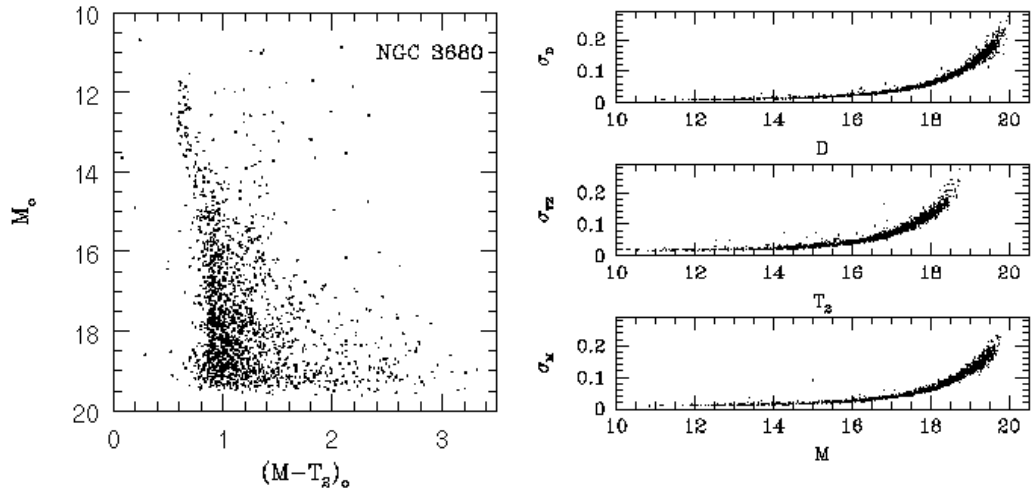}
\figcaption[5.ps]{Color-magnitude diagram for our
  catalogue of 1815 stars in the field of the open cluster NGC 3680.
  Right hand panels show the random errors in the photometry in our
  three filters.  Photometry is corrected under the assumption of
  $E(B-V) = 0.046$.}
\end{figure}

\begin{figure}
\includegraphics[angle=0,scale=1]{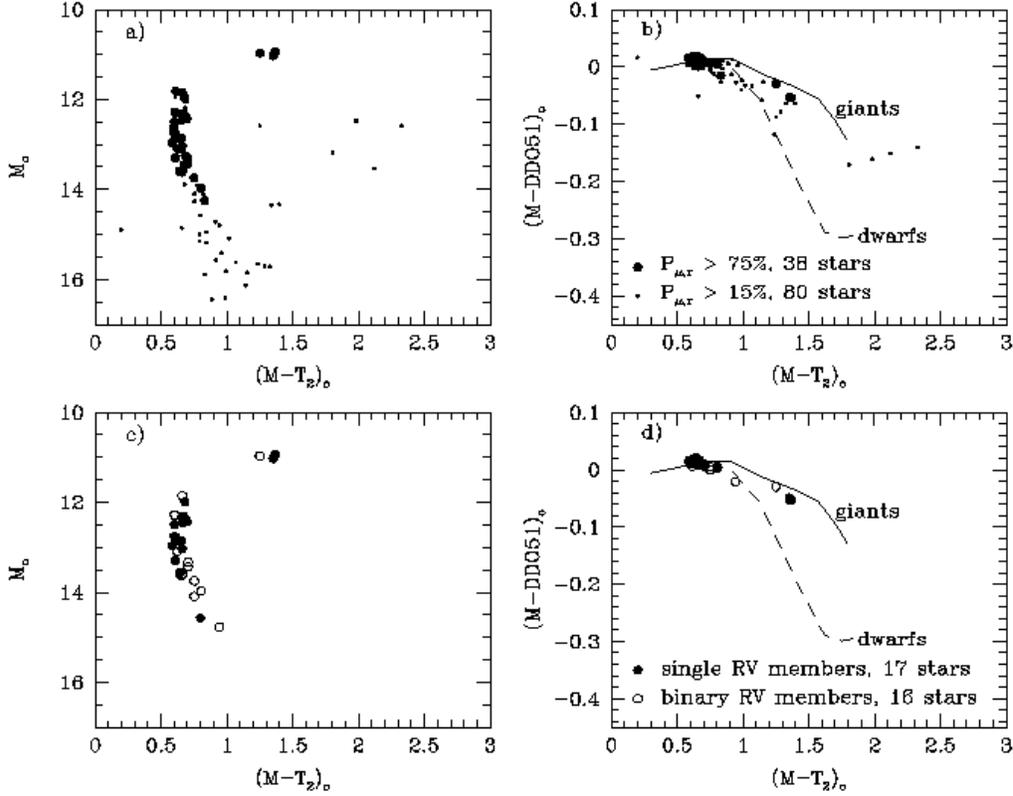}
\figcaption[6.ps]{(a) Color-magnitude diagram for stars in our NGC
  3680 catalogue also having proper motions by K95 most
  consistent with cluster membership.  {\it Large symbols} denote stars
  with a high probability of membership, as evaluated by the joint
  proper motion-radius index, $P_{\mu,r}$, in their survey, while the
  {\it small symbols} indicate stars with less certain membership.  (b)
  The ($M-T_2$)$_o$-($M-DDO51$)$_o$ diagram for the same sample of
  stars.  The \citeauthor*{pal94} lines for giants ({\it solid}) and dwarfs
  ({\it dashed}) of solar metallicity are also shown. (c) The $(M-T_2,
  M)$ CMD for stars in our NGC 3680 catalogue also having precision
  radial velocity membership as determined by \citet{nor97}.  {\it Open
    symbols} show radial velocity binary stars, whereas {\it closed
    symbols} show stars exhibiting no binarism in their radial
  velocities.  (d) The ($M-T_2$)$_o$-($M-DDO51$)$_o$ diagram for the
  radial velocity members of NGC 3680. The \citeauthor*{pal94} solar
  metallicity lines from panel (b) are again plotted.}
\end{figure}

\begin{figure}
\includegraphics[angle=0,scale=1]{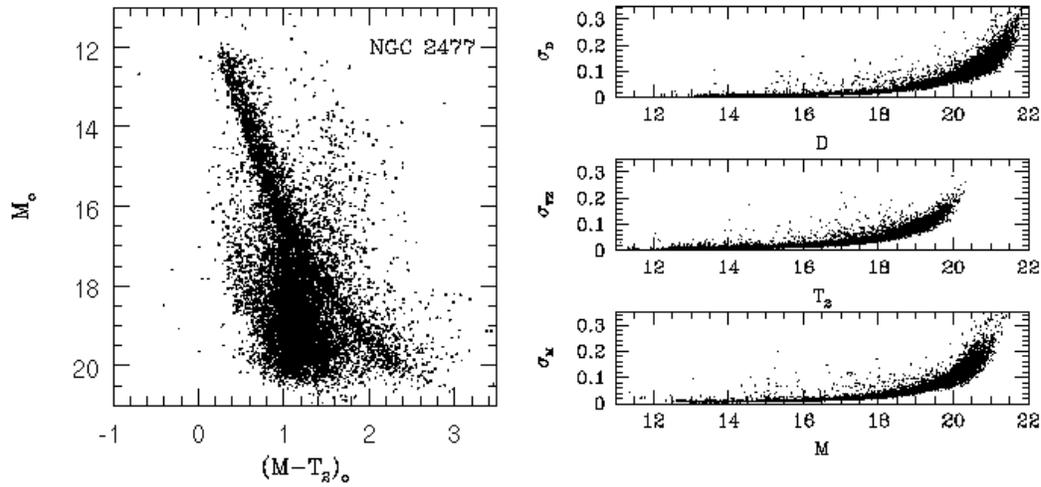}
\figcaption[7.ps]{Color-magnitude diagram for 11,300 stars
  in the field of the open cluster NGC 2477, with near solar abundance.
  The field shows substantial, and probably variable, reddening, from
  $E(B-V)=0.2$ to $0.4$.  The cluster is dereddened with the assumption of
  a mean $E(B-V)=0.33$.  The scattering of stars upward from the main
  sequence is likely due to differential reddening in the field, and
  this is supported by the lessening of the effect on the lower main
  sequence, which more nearly parallels the reddening vector.  The right
  hand panels show the random photometric errors in our catalogue.}
\end{figure}

\begin{figure}
\includegraphics[angle=0,scale=1]{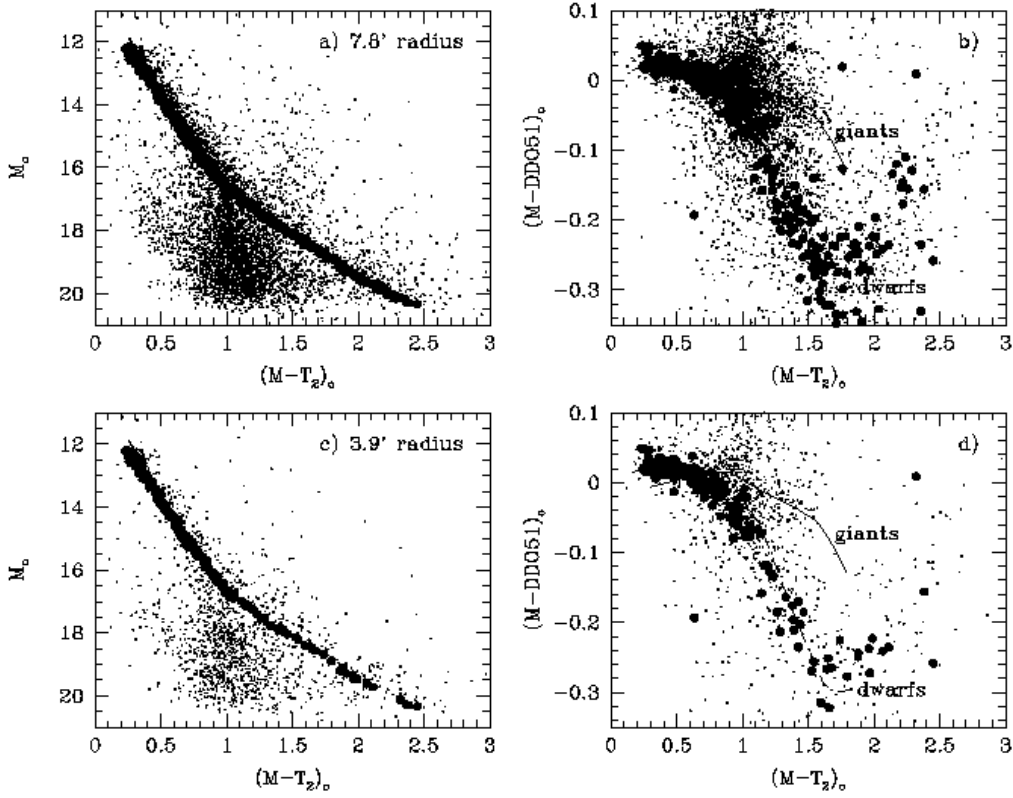}
\figcaption[8.ps]{(a) Color-magnitude diagram for a more limited
  field of radius $7\farcm8$ around the center of the open cluster NGC
  2477.  The {\it large symbols} are a ``by eye'' selection of the
  apparently least-reddened main sequence for the cluster.  Without
  ancillary membership information, some field star contamination of
  this latter sample might be expected.  The resultant
  ($M-T_2$)$_o$-($M-DDO51$)$_o$ distribution of the selected sample is
  shown in (b).  The \citeauthor*{pal94} lines for giants ({\it solid}) and
  dwarfs ({\it dashed}) of solar metallicity are also shown.  (c) An
  even more restricted cut in radius around the center of the cluster,
  with radius $3\farcm9$ around the cluster center.  Again, we select by
  eye a representative least-reddened main sequence for the cluster,
  shown with {\it large symbols}.  The stars in panel (c) are shown in
  the two-color distribution in (d).  The \citeauthor*{pal94} solar
  metallicity lines from panel (b) are again plotted in (d). The
  distributions in both panels (b) and (d) suggest that the ``by eye''
  selections are yielding mainly dwarf stars.}
\end{figure}

\begin{figure}
\includegraphics[angle=0,scale=1]{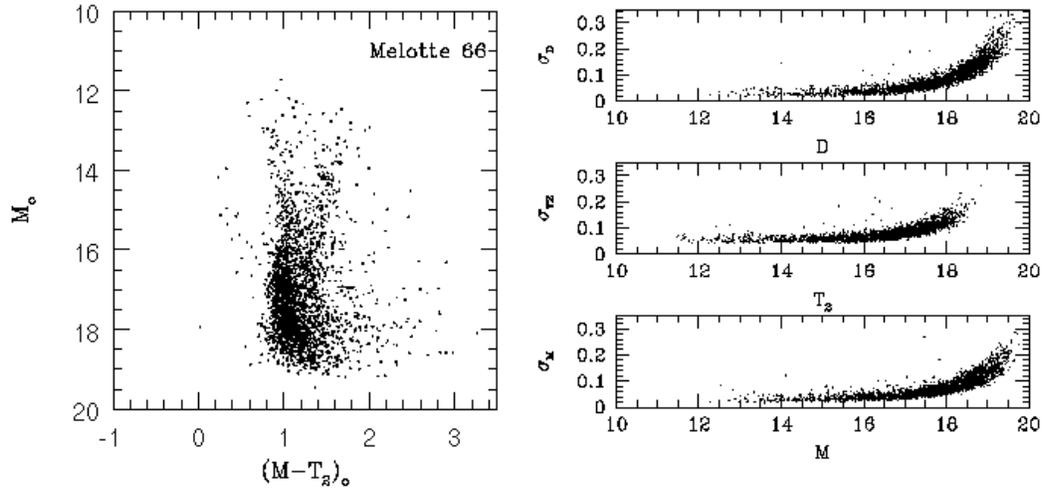}
\figcaption[9.ps]{Color-magnitude diagram for the open
  cluster Melotte 66.  Our catalogue contains 2643 stars from one
  telescope pointing.  We adopt a reddening of $E(B-V) = 0.16$ for the
  cluster.  The right hand panels give the random errors in the
  photometry.}
\end{figure}

\begin{figure}
\includegraphics[angle=0,scale=1]{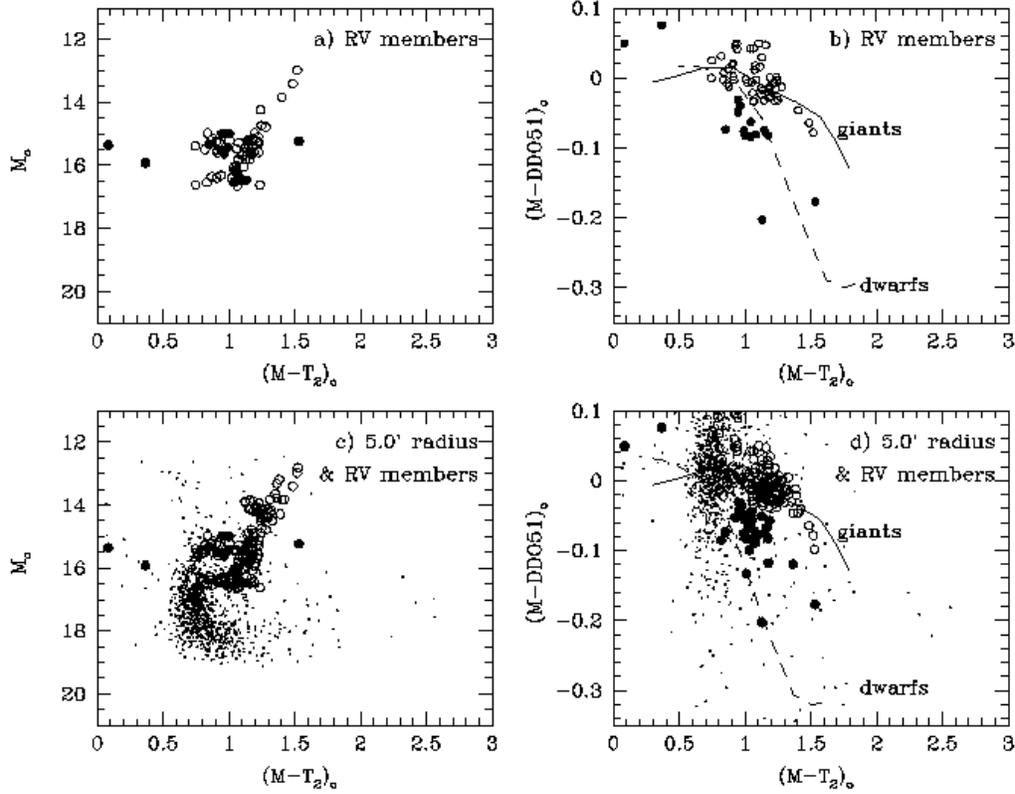}
\figcaption[10.ps]{(a) Color-magnitude diagram for radial velocity
  members of Melotte 66 as determined by the sources cited in the text.
  The double subgiant branch of the cluster, a result of a high binary
  fraction, can be seen.  The associated two-color distribution is given
  in (b), with the \citeauthor*{pal94} lines for giants ({\it solid}) and
  dwarfs ({\it dashed}) of solar metallicity also shown.  The {\it
    solid circles} mark stars that seem to fall in unexpected regions
  for giant stars in the two-color diagram.  A number of these seem to
  be members of the upper subgiant branch, and binarity may therefore be
  the cause for the observed deviation.  Many of these ``problem stars''
  also have abnormally large DAOPHOT $\chi$ values compared to other
  stars at the same magnitude.  To clarify the MSTO and subgiant region
  of Melotte 66 we show all stars within $5\farcm0$ of the cluster
  center in (c) and (d).  Then, using the binary and single star
  isochrones in \citet{kas97}, as well as the known radial velocity
  Melotte 66 members in panel (a) -- all of which are shown in (c) and
  (d) regardless of their radius from the cluster center -- as guides,
  we trace out the pair of giant/subgiant branches to their respective
  main sequence turnoff points, showing the selected additional stars as
  large symbols.  The {\it dots} are all remaining stars within
  $5\farcm0$ of the cluster center.  As in panels (a) and (b), we show
  ``problem giant/subgiant'' stars -- those with unexpected colors in
  the two-color diagram -- as {\it solid circles}.  The latter seem to
  be more associated with the binary sequence.  Panel (d) demonstrates
  how the two-color diagram becomes decreasingly effective as a surface
  gravity discriminator as the subgiant branch merges into the main
  sequence. The \citeauthor*{pal94} solar metallicity lines from panel
  (b) are again plotted in (d).}
\end{figure}

\begin{figure}
\includegraphics[angle=0,scale=1]{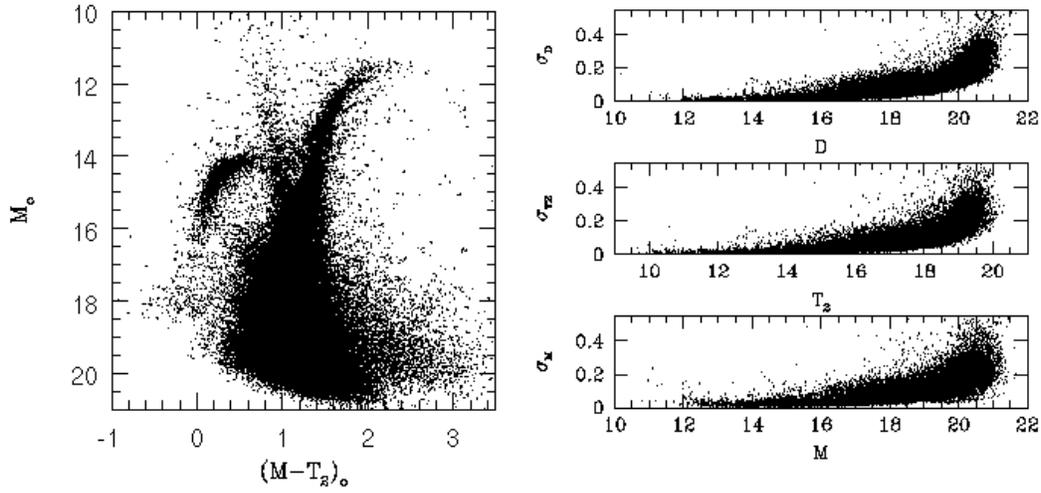}
\figcaption[11.ps]{$\omega$ Cen color-magnitude diagram for over
  100,000 stars in our catalogue generated from a $3\times3$ grid of
  pointings around the cluster, less the central, saturated core.  Some
  stars have been observed two or three times if they happened to fall
  in CCD frame overlap regions; these multiple detections have not been
  removed from the DAOPHOT catalogue or the diagram, which contain a
  quarter million entries.  Some of the very reddest, highest luminosity
  giants have been lost due to saturation on the (fairly short) CCD
  integrations. The right hand panels give the random errors in the
  photometry. The vertical thickness in the error plots, compared with
  Figures 5, 7 and 9, is due to the fact that the individual pointings
  which make up the $\omega$ Cen database are not all to the same
  depth.}
\end{figure}

\begin{figure}
\includegraphics[angle=0,scale=1]{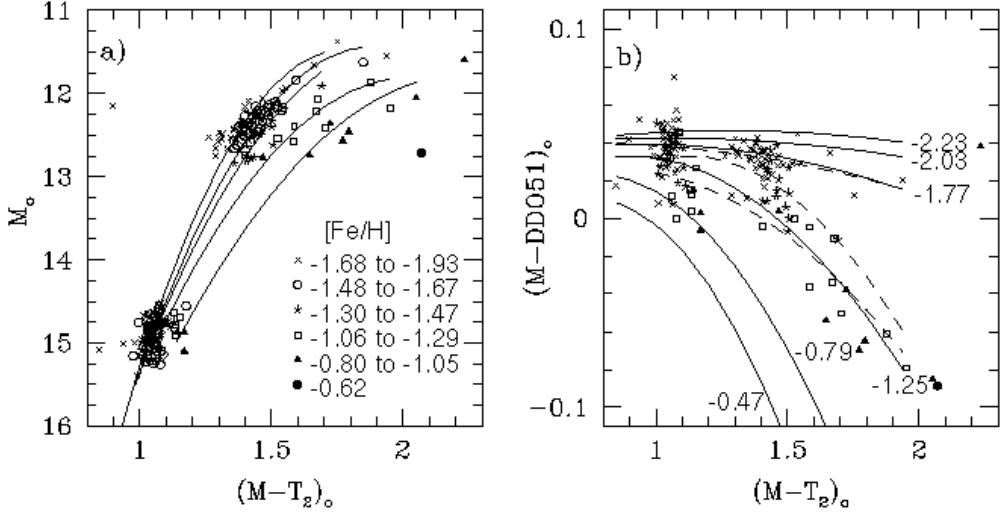}
\figcaption[12.ps]{(a) $\omega$ Cen color-magnitude diagram for
  giant stars with radial velocity and proper motion membership as well
  as [Fe/H] determinations from \citet{sun96}.  The latter chose stars
  from two magnitude ranges, one below and one above the horizontal
  branch.  The legend describes the abundance groupings used for the
  giant stars. Second order fits to the RGBs of the various metallicity
  groupings are overplotted.  The curves are in the order of abundance
  indicated in the panel legend, with abundance decreasing upward.  The
  associated distribution in the two-color diagram is illustrated in
  panel (b).  Second order fits to the two-color distributions for
  $\omega$ Cen giants of different metallicities, with the fits
  constrained by the color-color relations in \citeauthor*{pal94} (see text).
  The coefficients of the fits are given in Table 2. Note that the
  metallicities of the lines differ from those in panel (a) and Table 1;
  in panel (a) the intervals are set by the data from $\omega$ Cen
  alone, while in panel (b), we are constrained by the metallicity
  intervals chosen by \citeauthor*{pal94}.  The original empirical fits to the
  distribution of the (from top to bottom) [Fe/H]$=-1.74$, $-1.40$ and
  $-1.20$ groups are plotted in (b) (dashed lines) in order to
  illustrate the reasonable match obtained from the translated
  \citeauthor*{pal94} curves to the $\omega$ Cen data.}
\end{figure}

\begin{figure}
\includegraphics[angle=0,scale=1]{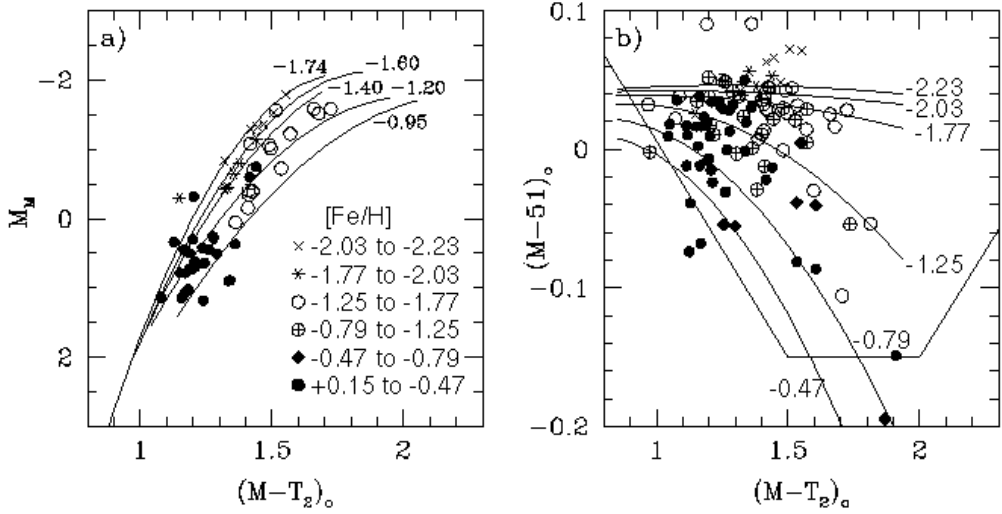}
\figcaption[13.ps]{Summary of open and globular cluster data from
  sources in Table 3.  (a) Color-magnitude diagram of open and globular
  cluster giants having $T_1$ magnitudes available from the Geisler
  collaboration for conversion to $M$ magnitudes.  Distance moduli were
  adopted as given in Table 3, and we have assumed $R_V=A_V/E(B-V)=3.1$
  and the reddening relations in \S 2.3. The iso-metallicity lines are
  from Figure 12a. (b) The resulting two-color diagram of the cluster
  giants from the data in Table 3, with the iso-metallicity lines from
  Figure 12b.  For both panels (a) and (b) giant stars are grouped into
  metallicity bins indicated in panel (a). Note that the metallicity of
  the lines shown in panel (a) and panel (b) differ, as discussed in the
  caption for the previous figure.  The {\it straight solid lines} in
  (b) are taken from those shown in Figure 14 below.}
\end{figure}

\begin{figure}
\includegraphics[angle=0,scale=1]{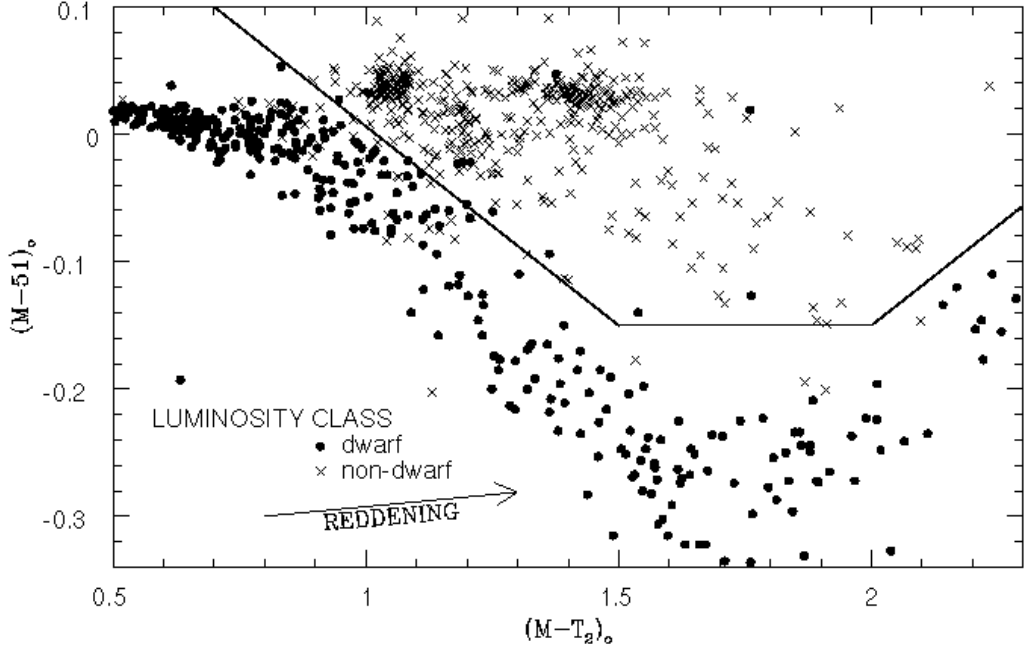}
\figcaption[14.ps]{Summary ($M-T_2$)$_o$-($M-DDO51$)$_o$ diagram of
  all giant and dwarf data discussed in this paper.  The {\it crosses}
  represent giant stars presented in Figure 13, our red giants from NGC
  3680 (both single and binary radial velocity members in Figure 6d) and
  $\omega$ Cen (Figure 12b), the red giant and subgiant branch of
  Melotte 66 (radial velocity members from the single star locus only,
  Figure 10b), the solar abundance field giants from \citet{gcm91}, as
  well as all evolved stars from Geisler's (1984) classes ``IV-V'' or
  class I.  The {\it solid circles} (dwarfs) are \citet{gei84}
  luminosity class V and V-VI stars, and our open cluster main sequences
  from NGC 3680 (both single and binary radial velocity members in
  Figure 6d) and NGC 2477 (those stars selected as main sequence in
  Figure 8d).  Giant stars ({\it crosses}) are, for most ($M-T_2$)$_o$
  colors, well separated from the dwarfs, for all metallicities.  The
  {\it straight solid lines} roughly demarcate the region above which
  one might expect to find predominantly subgiant and giant stars.}
\end{figure}

\end{document}